%% file: Gonzalez.tex
\documentclass[preprint]{aastex}

\usepackage{epsfig,lscape}

\slugcomment{To appear in AJ, NN.}
\shorttitle{The Opacity of Galaxies}
\shortauthors{Rosa Gonz\'alez et al.}

\begin{document}

\newcommand{\hi}{H\,{\sc i}}
\newcommand{\hii}{H\,{\sc ii}}

\title{The Opacity of Nearby Galaxies \\
from Counts of Background Galaxies: \\[0.2in]
II. Limits of the Synthetic Field Method\altaffilmark{1}}

\author{Rosa A.\ Gonz\'alez\altaffilmark{2,3} and Laurent 
Loinard\altaffilmark{4}}\affil{Instituto de Astronom\'{\i}a, UNAM, Unidad 
Morelia, Michoac\'an, M\'exico, C.P. 58190}

\author{Ronald J. Allen\altaffilmark{5}}
\affil{Space Telescope Science Institute, Baltimore, MD 21218}

\and

\author{S\'ebastien Muller\altaffilmark{6}}
\affil{Institut de Radio Astronomie Millim\'etrique, F-38406 St. Martin 
d'Heres, France}

\altaffiltext{1}
{Based on observations with the NASA/ESA Hubble Space Telescope obtained at 
the Space Telescope Science Institute, which is operated by the Association 
of Universities for Research in Astronomy, Incorporated, under NASA contract 
NAS5-26555.}
\altaffiltext{2}{E-mail address: {\tt r.gonzalez@astrosmo.unam.mx}}
\altaffiltext{3}{Previously: National Science Foundation International 
Research Fellow at the Observatoire de Gen\`eve, CH-1290 Sauverny, 
Switzerland} 
\altaffiltext{4}{E-mail address: {\tt l.loinard@astrosmo.unam.mx}}
\altaffiltext{5}{E-mail address: {\tt rjallen@stsci.edu}}
\altaffiltext{6}{E-mail address: {\tt muller@iram.fr}}

\newpage

\begin{abstract}
Recently, we have developed and calibrated the Synthetic Field Method to 
derive the total extinction through disk galaxies. The method is based on 
the number counts and colors of distant background field galaxies that can 
be seen through the foreground object; it is the {\it only} method capable 
of determining extinction without {\it a priori} assumptions about the dust 
properties or its spatial distribution, and has been successfully applied 
to NGC 4536 and NGC 3664, two late--type galaxies located, respectively, at 
16 and 11 Mpc.

Here, we study the applicability of the Synthetic Field Method to $HST$ images 
of galaxies in the Local Group, and show that background galaxies cannot be 
easily identified through these nearby objects, even with the best resolution 
available today. In the case of M~31, each pixel in the $HST$ images contains 
fifty to one hundred stars, and the background galaxies cannot be seen because 
of the intrinsic granularity due to strong surface brightness fluctuations. In 
the LMC, on the other hand, there is only about one star every six linear 
pixels, and the lack of detectable background galaxies results from a 
``secondary'' granularity, introduced by structure in the wings of the point 
spread function.

The success of the Synthetic Field Method in NGC 4536 and NGC 3664 is
a natural consequence of the reduction of the intensity of surface
brightness fluctuations with distance.  When the dominant confusion
factor is structure in the PSF wings, as is the case of $HST$ images
of the LMC, and would be the case of M~31 images obtained with a 10-m
diffraction-limited optical telescope, it becomes {\it in principle}
possible to improve the detectability of background galaxies by
subtracting the stars in the foreground object. However, a much better
characterization of optical PSFs than is currently available would be
required for an adequate subtraction of the wings. Given the
importance of determining the dust content of Local Group galaxies,
efforts should be made in that direction.

\end{abstract}

\keywords{galaxies: general --- galaxies:individual (M~31) --- 
galaxies:individual (NGC~4536) --- galaxies:individual (LMC) ---
galaxies:ISM --- galaxies:Local Group --- galaxies:spiral --- 
galaxies:statistics --- ISM:dust,extinction --- ISM:general}

\section{Introduction} \label{intro}

\subsection{Some history} \label{history}

How much dust is there in galaxy disks, and where is it? Reliable answers 
to these questions have remained elusive, although they bear directly on 
our understanding of many astronomical problems. Dust has to be properly 
accounted for to infer luminosity, stellar composition, and mass distribution 
correctly. Dust could be veiling correlations between global properties of 
galaxies and affecting our view of their dynamical evolution. It could be 
hiding significant amounts of luminous baryonic matter, or could trace 
extremely cold gas, and hence have implications for the dark matter problem.
Dust in disks and halos of foreground spirals could be distorting our 
perception of the high-redshift universe: it could be partly responsible for 
the apparent decrease of quasi stellar objects (QSOs) at high redshifts (e.g., 
\citealp{fall93}), and for the problems encountered in detecting primeval 
galaxies. And dust in the precursors of present day galaxies could completely 
change our most recent picture of the star formation history of the universe 
\citep{mada96, meur97}. To give just one more example, the Tully-Fisher 
correlation as a measure of cosmological distance depends on inclination 
corrections which could be significantly sensitive to dust.

Much work has been devoted to the problem of dust in galaxies, but different 
authors have often come up with widely differing results, from disks that 
appear significantly opaque to those basically transparent, and everything in 
between (see \citealp{davi95} and references therein). The virtues and 
disadvantages of different tests of opacity of spiral disks have been 
discussed at length in Gonz\'alez et al.\ (1998, Paper I), and will not be 
repeated here. In summary, only one method is sensitive to the total 
amount of dust in disks, independently of its temperature, and
does not require that any assumptions be made, neither about the geometry of 
the background probes, nor about the distribution of the absorbing
dust (e.g., uniform screen or a collection of opaque blobs): 
the comparison of counts of background field galaxies seen 
through the disk of the foreground galaxy to counts of galaxies in reference 
fields. 

This technique was pioneered by \citet{shap51}. Before Paper I, however, it 
had only been applied to the Magellanic Clouds \citep{shap51, wess61, hodg74, 
gurw90}. Furthermore, \citet{wess61} deemed his results as ``inaccurate'', 
while \citet{gurw90} qualified theirs as ``tentative.'' The conceptually very 
simple idea of galaxy counts is, unfortunately, very difficult 
indeed to implement in practice. The general distribution of disk stars in the 
foreground galaxy reduces the contrast of background galaxy images, and 
foreground star clusters and \hii\ regions add strongly to the confusion. 
A first attempt to measure this effect was made by \citet{macg75}, who used a 
plot of galaxy counts vs.\ star counts in the Small Magellanic Cloud to 
subtract the effects of ``masking by stars'' from those of extinction. 
In Paper I, we developed the ``Synthetic Field Method'' (SFM) in order to 
quantify and calibrate the effects of crowding and confusion, and to 
determine the accuracy with which statements can be made about the total 
opacity and reddening of an average line of sight through a foreground galaxy. 
We then applied it to NGC~4536, a bright spiral galaxy in the direction of 
Virgo, at a distance of 16.2 Mpc \citep{saha96a}, and to NGC~3446, a small 
Magellanic irregular at a distance of  11.9 Mpc. In the SFM, images of a 
suitable reference field are added directly to the images of the foreground
galaxy under study. The extinction through that foreground galaxy is then
determined by comparing the numbers of ``real'' background galaxies, not to 
the number of galaxies in the original reference field image, but to the number
of galaxies in the reference field that can still be identified after its
addition to the foreground galaxy image. This procedure was hinted at by 
\citet{wess61}, and by \citet{gurw90}; they superimposed negatives of fields 
with and without foreground galaxy as an experiment to estimate the reduction 
in galaxy counts owing to crowding and confusion, but unfortunately did not 
correct their results accordingly.

The advent of the Hubble Space Telescope ({\sl HST}) has made the method more 
easily applicable, thanks to its superb resolution and sensitivity, as well 
as to the stability of the sky, the photometry, and the point spread function 
(PSF). Other factors that favor the application of the method nowadays are the 
availability of deep and medium deep {\sl HST} images that can be used as 
reference fields \citep*[Williams et al.\ 1996; Griffiths et al.\ 1994; 
Ostrander et al.\ 1998][]{ratn99a,ratn99b}, and the access to automated 
surface photometry packages \citep[for example,][]{bert96}, which allow an 
accurate assessment of galaxy detection limits.

De Vaucouleurs (1995) suggested that, after the Magellanic Clouds, the best 
targets would be the next largest galaxies in angular size, i.e., M~31, M~33, 
M~51, M~81, M~101. Of course, the surface density of background field galaxies 
is roughly constant in any direction of the sky, but closer galaxies will 
cover a larger angular area, allowing for results with better spatial 
resolution. Indeed, M~31 appears as an excellent potential candidate to apply 
the SFM because it is so nearby, and so well studied. Not only it could offer 
a unique opportunity to study the distribution of dust on small scales, and in 
particular to investigate the differences between arm and interarm regions, 
and between different galactic radii. But also, large--scale, high--resolution 
maps of the 21--cm line of \hi\ \citep{brin84}, and of the 2.7--mm line of CO
\citep{loin99,niet2000} have been obtained, as well as large--scale maps of 
the distribution of warm dust (Haas et al. 1998, and references therein). Once 
the total dust content is known, it would become possible to determine 
unambiguously whether there exists a significant fraction of very cold dust 
(at T $<$ 10 K) -- as has been suggested in the past -- that has so far 
escaped detection in the far--infrared. Also, assuming a canonical 
dust--to--gas ratio, the total gas content could be measured by means 
completely independent from the usual methods. In particular, the mass of 
molecular gas could be estimated without having to rely on the (highly 
uncertain -- see Loinard \& Allen 1998) CO to H$_2$ ``conversion factor''. 

\subsection{It is much harder than it looks} \label{motiv}

In a nutshell, the motivation for this paper is the fact that -- as we shall
see momentarily -- applying the SFM to {\sl HST} Wide Field Planetary Camera 
2 (WFPC2; \citealp{holt95}) data of M~31 or the LMC has not proven easy or 
very fruitful. Almost no background galaxies were visible through the disk of 
M~31, and very few could be seen through the Magellanic Clouds (see below). In 
the case of the LMC, we were particularly surprised because a very simple 
calculation (dividing 10$^{10}$ stars by the area of the galaxy or $\sim$ 68 
kpc$^2$) shows that there is a star only every $\sim$ 4 linear Wide Field 
Camera (WFC) pixels; i.e., with 0\farcs1 pixels, we should already be looking 
between the stars!\footnote{M~31 is about 14 times further and has more stars 
per unit area (see \S\ref{starsim}) than the LMC, so the WFC Andromeda data 
should not have ``empty'' pixels. But Virgo galaxies are more than 20 times 
further away than M~31 and hence have many more stars per pixel, and yet the 
SFM was successfully applied there.} Moreover, since it is also very difficult 
to see any reference galaxies in the synthetic combined images, it is 
impossible to tell whether the paucity of ``real'' background galaxies 
is caused by extinction or by lack of contrast against the galaxy foreground. 
In order to illustrate the problem,
figure \ref{lmc_color} shows color cutouts of one of the 
WFPC2 subfields of the LMC we have used, without ({\it left}) and
with ({\it right}) the HDF-N added in;  
finding galaxies, especially the simulated ones, behind NGC~4536 in Virgo, in 
data with similar surface brightness, had been comparatively much easier -- 
albeit easy, it was not.

In summary, while the SFM provides useful results at the distance of Virgo, 
it seems to fail in nearby objects. The purpose of the present paper is to 
determine quantitatively the conditions in which the SFM can yield useful 
results, and whether we can expect (or when we can expect) to use it to 
determine accurately the total opacity of Local Group and nearby galaxies.
To this end, we will first look into the characteristics, both of the 
NGC~4536 images where the SFM was applied successfully, and of 
those images where it failed. Subsequently, we will perform simulations at 
different levels. First, we will artificially reproduce the stellar fields of 
our selected images; we will constrain these simulations with published 
luminosity functions, as well as with the stellar counts, surface brightnesses,
and pixel--to--pixel dispersions of the data. Next, we will simulate either 
changes of surface brightness at fixed resolution, or changes of resolution 
(to mimic improvements that can be expected with increasingly bigger, 
diffraction--limited telescopes). To each simulated image with different 
brightness/resolution, we will add the HDF-N (used as reference field), in 
order to estimate the number of background galaxies that we can expect to see 
in the absence of extinction.

\section{The observational material} \label{data}

All the actual data used are archival {\sl HST} WFPC2 images. Table 1 lists 
the coordinates of the targets, and the total exposure times and observation 
bands. With the exception of the HDF-N, the data were retrieved from the 
archive after pipeline calibration, and subsequently cleaned of ``warm'' 
pixels and cosmic rays, aligned, and coadded. A full description of these 
steps for the NGC~4536 data has been given in Paper I; there are two small 
differences for the additional data used here. First, cosmic rays in the 
M~31--Inner field and in the LMC data were cleaned by rejecting very high 
counts in pixel stacks \citep{voit97}, and not by anti-coincidence in CR-split 
exposures \citep*{ande95}. One of the 3 images in each of the two (F555W and 
F814W) M~31--Inner mosaics has a much shorter exposure time (300 and 200 s, 
respectively, in $V$ and $I$). Regarding the LMC, there were only two images 
available in each filter (F606W and F814W); we made a stack of the 4 images to 
identify the cosmic rays. The second difference in the reduction is that, for 
the Local Group data, no relative shift between individual images had to be 
found, since all the exposures for each field were taken, respectively, on the 
same epoch and with the same pointing.

\subsection{HDF-N} \label {dhdf}

We used the Version 2 Release WFC images of the Hubble Deep
Field (HDF-N; \citealp{will96}). There are 3 contiguous WFC fields in this 
data set (and one PC field which we do not use here); we refer to them 
as WF2, WF3, and WF4. Each WFC field was 
recorded in the 4 photometric bands designated F300W, F450W, F606W, and F814W
(cf. \citealp{voit97}); we have not used the images in the UV band F300W. Two 
processing steps, described at large in Paper I, were carried out on the HDF-N 
data set in order to render it compatible with the other archival WFPC2 images 
used here. Briefly, the HDF-N images were first re-binned to their original 
pixel size of 0\farcs1; second, since there is no HDF-N image in the F555W 
passband, we created one by interpolating linearly between the existing F450W 
and F606W images, taking into account the differing responses of the three 
filters. These ``interpolated'' F555W image was not needed for the LMC
simulations, given that the LMC data were taken through the F606W filter.

\subsection{NGC~4536} \label{dn4536}

The images of NGC~4536 were originally acquired with the WFPC2 for the 
calibration of type Ia supernovae in nearby galaxies as standard candles
\citep{saha96b}; they have already been published by \citet{saha96a}, who give 
a very detailed account of the observations and the pipeline calibration 
(i.e., bias- and dark- subtraction, and flat-fielding) of the data.

There are thirty--four 2000 s individual exposures of NGC~4536 in the F555W passband, 
taken at 17 discrete epochs between 1994 June 3 and 1994 August 9; in the 
F814W passband, ten 2000 s images were taken at 5 discrete epochs, over the 
same two-month period. We cleaned the pipeline-calibrated frames of 
``warm pixels'' and cosmic-rays, aligned them, and combined all the 
data in each passband into a single 
deep image; the resulting mosaics have total exposure times of, respectively, 
$6.8 \times 10^4$ seconds in the F555W passband, and $2 \times 10^4$ seconds 
in the F814W passband.

\subsection{M~31 and the LMC}\label{m31}

Two fields were considered in M~31. The data on the ``Inner'' field 
(M~31--Inner) were obtained on 1997 August 14 for a program on stellar 
populations in the M~31 disk (Trauger, Holtzman, \& Gallagher 1997). The 
total exposure times are 2400 s at F555W and 2500 s at F814W. The field is 
located at about 5 kpc from the center of M~31. It was chosen specifically 
to avoid spiral arms and to have low, uniform extinction; hence, it shows 
very little structure. The data on the ``D478'' field of M~31 (M~31--D478) 
were obtained in parallel mode on 1996 July 8, as part of the Medium Deep 
Survey \citep{grif94,ostr98,ratn99a,ratn99b}. The total exposure times are 
5600 s at F555W and 6400 s at F814W. We retrieved them because the field 
overlaps part of the dark dust cloud D478, which is located at 2.5 kpc from the 
center of the galaxy \citep{hodg81}, and has been extensively studied by Allen and 
co--workers \citep[Allen \& Lequeux 1993, 1994;][]{loin98}.

The data on the LMC were obtained in parallel mode on 1996 January 30, also 
as part of the Medium Deep Survey. The field is located $\sim$ 1\fdg3 from 
the center of the LMC, and these same observations have been amply discussed 
by \citet*{elso97}.

\section{Diagnostic of the problem} \label{problem}

Before even trying to test for the presence of extinction in the M~31 and LMC
fields, we added to each of the WFC subimages each of the 3 HDF-N WFC chips. 
These ``baseline'' simulations, where the 3 chips of the control field are 
added one at a time without attenuation, establish the capability of the 
method to discriminate between crowding and extinction (Paper I).

In the case of the M~31--D478 field, regardless of what WFC subimage we 
analyzed, we could barely see the one or two brightest galaxies of each of 
the HDF-N chips. We could also see at most one or two ``real'' background 
galaxies in the whole WFC field. In the M~31--Inner field, we could see an 
average of 7 HDF-N galaxies and 4 ``real'' galaxies per WFC chip. As for the 
LMC, the simulations yielded an average of 22 galaxies per chip, and we 
could count 10 ``real'' background galaxies per subimage, on average.
Assuming that the error in the galaxy counts equals $\pm$3.5$\times N^{1/2}$ 
(see \S \ref{results}), these numbers translate roughly into $A_V = 
1.2\pm3.1$ mag for the M~31-D478 field; $A_V = 0.3\pm1.3$ mag for the
M~31-Inner images; and $A_V = 0.9\pm1.1$ mag for the LMC. 
Given the errors, 
these are not tremendously useful results! 
For comparison, in similar ``baseline'' simulations of NGC~4536 (Paper I), 
we found an average of 28 HDF-N galaxies in each of the 2 subimages that 
comprised the arm region (WFC2 + WFC3), and of 38 control galaxies in 
the interarm (WF4) region; the corresponding uncertainties on the opacity were
0.2--0.3 mag.

We hypothesized early that image ``granularity'' might have an important 
role in the success or failure of the SFM; hence, we 
searched for diagnostics of such image granularity.
Average means or standard deviations 
calculated across entire images would not work,
because ``smooth'' data with large-scale spatial gradients and 
``flat'' data with large pixel-to-pixel differences 
would produce similar values (see Figure \ref{rmsfig}). Consequently, 
we considered instead the distributions of pixel 
means and rms variations in squares 100 pixels on the 
side. We compared the
different distributions shown by the
M~31, LMC, and NGC~4536 data; they are
listed in Table 2.
While the means reflect the surface brightnesses, the standard 
deviations constitute a diagnostic of the image granularity. For example, 
very broad distributions of mean values point to large-scale structure, due,
for instance, to stellar associations and dust lanes.

The case of the M~31--D478 field is clearcut. It is on average more than two 
times brighter than NGC~4536 (Table 2). More significantly, it is much more 
``granular'', as shown by a mean rms 5 times larger, relative to the mean 
surface brightness. Large scale structure does not seem to contribute to the 
problem: the FWHM of the distribution of surface brightnesses in D478 is 
$\sim$ 14\%, compared to the mean; this is of the same order as the spread in 
the interarm field of NGC~4536 (12\%), the subimage with the least large-scale 
structure. The M~31--Inner field is indeed quite structureless, as attested 
by a distribution of surface brightnesses with an average FWHM of less than 
1\%! It is barely 10\% brighter than NGC~4536 but, on the other hand, it is 
again 5 times more ``granular.''

Finally, the LMC field is hardly brighter than NGC~4536; with the exception of 
the chip WF4, the structure shown by the data is also of the same order. 
However, the pixel-to-pixel variations are larger than 300\% of the mean. 
Such a large value of the rms is, indeed, a
sign that we are looking between the stars. 

\section{Simulations of stellar foreground} \label{starsim}

To quantify the effect of the granularity,
we mimicked the various observations used with 
artificial stellar fields, and measured their effect on the  
detectability of background galaxies. These simulations were all produced in the $V$ 
passbands, i.e., as if they had been obtained through the {\sl HST} filters F555W or F606W. In 
general, colors are crucial for the identification of background galaxies and 
hence for the success of the SMF. However, the present work is centered on 
simulations of stellar populations; population models, as well as 
magnitudes for different stellar types, are most readily available at $V$. 

The simulations had several levels: first, we reproduced the LMC and 
M~31--Inner {\sl HST} WFPC2 data, guided by published luminosity functions 
and/or the star counts of the data itself, and constrained by the mean and 
rms of the images. We chose to use the M~31--Inner field because of its 
,``intermediate'' surface brightness and because, as stated above (\S 
\ref{data}), it has very little large-scale structure and is dominated by 
image granularity. It has the additional advantage that it has roughly the 
same surface brightness as the NGC~4536 data (Table 1), so we could also use 
it as a basis to simulate a galaxy disk at the distance of Virgo
(see below \S \ref{virgosim}. In the case 
of the LMC, we really wanted to know why we could not detect more background 
galaxies in between the stars. At this stage, we always used a Tiny Tim 
\citep{kris01} PSF. 

The second step was to simulate the change in resolution offered, first by 
the smaller pixels of the Advanced Camera for Surveys (ACS), which has just been
put on-board the {\sl HST}, then by increasingly bigger, 
diffraction--limited telescopes. This was achieved by changing the number 
of stars per pixel. Here, we did two subclasses of simulations. In one case, 
we used Gaussian PSFs, and assumed that they were fully sampled. In the other, 
we used the Tiny Tim PSFs foreseen for the two cameras (Wide Field, ACS-WFC, 
and High Resolution, ACS-HR) of the ACS. Since the PSF for the ACS-HR should 
be Nyquist-sampled in the visible, we used its PSF to simulate the 
observations with bigger telescopes.

For completeness, we have also done a slightly different kind of simulation, 
which will have immediate applicability to predict the potential success of 
the SFM in existing archival images. We have mimicked changes in surface 
brightness with radius of the disk of M~31, at the resolution and with the 
PSF of the WFPC2 on board {\sl HST}. Assuming a constant stellar population, 
this also boils down to changing the number of stars per pixel. 
Finally, for every object, and at each different brightness/resolution, we 
have added the HDF-N to the data, in order to assess the number of background 
galaxies that we can expect to see in the absence of extinction. 

The background galaxies have been counted by eye. We have performed 
multiple checks since we started working on the subject and, 
as long as it is the same
person who identifies the ``real'' and the control galaxies, this procedure
yields robust results.  
Eventually, it would be desirable to 
use an automated photometry package 
especially adapted to this kind of study; however, 
automatic star/galaxy separation in crowded fields is quite difficult.
For example, we have tested SExtractor \citep{bert96} on the F606W the WF3 field of 
the LMC (\S \ref{m31}), both with and without the HDF-N added in (see 
Fig.\ \ref{lmc_color}). The program finds
significantly more extended objects than we do galaxies by eye,
but it is unclear which of the identified structures are actual     
galaxies, since SExtractor finds more (185 vs.\ 153) of these         
extended objects in the image {\it without} the HDF-N! 
In the case of our simulations,
which do not include dust, HII regions, etc., star/galaxy separation
should be in principle easier than in real life; 
but in any case we considered this effort as unnecessary and 
beyond the scope of the present work. 
Lastly, we do not investigate the application of the SFM to    
hypothetical images produced by  
a large IR telescope in space, like the New Generation Space 
Telescope (NGST) is anticipated to be. This would have necessitated 
simulations not only of the stellar foreground, but also of the background 
galaxies.

We present results for infinite exposure times (noiseless data), and also for
``realistic'' integration times, i.e., 1800 s or 7200 s with 
the {\sl HST}. For telescopes with a hypothetical aperture D in meters, these 
times should be scaled by (2.4/D)$^2$.  
The area of sky considered is always the same for all examples in the paper, 
and consequently the clustering error (\S \ref{results}) has always 
been assigned in the same way. 
\footnote{We assume that the reduced field--of--view 
of bigger telescopes (which, 
in the absence of special
optics/detectors, is also proportional to D$^{-2}$) will
be compensated by the smaller exposure time required to achieve the same 
depth, and that it will be possible to make mosaics with the same area in
about the same total exposure time.}

\subsection{M~31--Inner} \label{m31simu}
  
We used stellar population models, and the constraints offered by the data, 
to determine an adequate mixture of stars. The bulk of them were taken to 
belong to an old population with a main-sequence (MS) upper-cutoff type F5V and 
0.05\% by number of K giants \citep{mccl68}.
To account for the fact that we were trying to simulate a disk, while 
\citet{mccl68} studied the bulge of the galaxy, we then added 2 \% by mass of 
a younger population (for this paper, percentage by mass 
is equivalent to percentage by number, because upper mass limits of different 
MS components are not so discrepant, and massive stars contribute 
very little to the total stellar mass). We have adopted this number for 
the contribution by young stars following our fit to the
arm and interarm region populations of M~99 \citep{gonz96}.\footnote{
There is also a hand--waving argument: if a galaxy has lived about 10 Gyr, 
and the spiral pattern goes around every 10$^8$ years, and the passage of the
arm induces star formation (which is of course debatable and debated; e.g.,  
\citealp{elme87}), the 
youngest population would be of the order of 1 \%.} 
In these and all subsequent cases, main-sequences go down to M6 stars, or about the 
hydrogen-burning limit. 
At this point, the two free parameters 
left were the upper-cutoff stellar type of the young population, and the 
number of stars per pixel of the old population. The combination that best fit 
the mean and rms of the data was an upper cutoff type B5V, and 50 old stars 
pixel$^{-1}$; the red giants and the young stars scaled accordingly. Stellar 
population synthesis models \citep{char91,bruz93} with the described 
composition will have a mass-to-light (M/L) ratio of 1.8 which, assuming a 
Salpeter initial mass function (IMF), will yield about 100 stars per pixel. 
However, the factor of two discrepancy could very well be explained by a 
different percentage of young population, a different IMF, a different 
foreground Milky Way extinction towards M~31 (which is hard to measure 
precisely because of the dust and gas in the Andromeda galaxy itself), 
possible error in the distance to Andromeda, and by internal extinction in 
M~31 (which, of course, is what we are trying to measure). We also note here 
that including more luminous stars (B2V, say) in the young population makes 
it possible to match at the same time the surface brightness of the data and 
the number of stars per pixel required by the M/L ratio of the stellar 
population synthesis models; however, it is questionable that we see B2V stars 
in the data, and the inclusion of these brighter stars increased the standard 
deviation of the simulations much beyond that of the data.
   
Salpeter luminosity functions were used for the main-sequences, and a power-law
of the magnitude with an exponent of 0.7 for the red giants
\citep{iben68}. A uniform spatial density was used for all three 
components. The lower and upper absolute magnitudes for all components of the 
stellar population are given in Table 3. We also show the number of stars per 
pixel, in each component, for our simulation of the M~31--Inner field at the 
resolution of the WFPC2; the relative contributions of the three components 
remain fixed at all angular resolutions. 

Figure \ref{cmfig} shows the V vs.\ V-I color--magnitude diagram (CMD) for the stars in 
the real M~31--Inner data. Stellar cores were fitted with DoPhot 
\citep*{sche93}, zero-points were found iteratively as prescribed by 
\citet{holt95}, and 
aperture corrections were applied (-0.62 at V, -0.77 at I). No reddening 
correction was performed. The box delineates the approximate area of the CMD 
populated by B stars for a distance modulus to M~31 of (M-m)$_o$ = 24.47 
\citep{stan98}; the rest of the stars in the diagram are either
red giants or asymptotic giant branch (AGB) stars. 
Assuming a Galactic reddening towards M~31 of E(B-V) = 0.062 
\citep*{schl98}; an internal reddening of E(B-V) = 0.05, or a third 
\citep{char00} of that found by \citet{mass95} for OB associations in 
Andromeda; and R $\equiv A_V$ / E(B-V) = 3.1, B5 MS stars 
($M_V$ = -1.1) will have $m_V$ = 23.7 in this diagram. (B2V stars, $M_V$ = 
-3, will have $m_V$ = 21.8.) The incompleteness in the diagram prevents us 
from using it to improve the estimate of the percentage of young population.

The positions and magnitudes for the artificial stars needed for the 
simulation were generated with the task {\it starlist}, within the software 
package IRAF\footnote{IRAF is distributed by the National Optical Astronomy 
Observatories, which are operated by the Association of Universities for 
Research in Astronomy, Inc., under cooperative agreement with the National 
Science Foundation.} \citep{tody93}. Computer memory limited the number of
stars that could be generated in each run of {\it starlist} to about 5x10$^5$.
This is sufficient to produce a square of 100 pixels on the side for the 
simulation of the M~31--Inner field with the resolution of the WFPC2. 

Next, the task {\it mkobjects}, also within IRAF, was used to produce an actual
image of the simulated stellar field out of the list of stars given by {\it 
starlist}. With Tiny Tim \citep{kris01}, we generated 
a PSF appropriate for a G5V star in the F555W passband, at the center
of chip WF3; 
we used only one PSF for all stars, regardless of their
stellar types or their positions in the image. 
We produced only one simulated 100x100 square to reproduce the M~31--Inner data; 
in order to create a frame with the angular size of the WFC 
subimages (80 \arcsec on the side), we copied the artificial stellar field 
many times over. This means that the 
simulation with the resolution of the WFPC2 reproduces only the means, and not 
the spreads, of the distributions of surface brightnesses and standard 
deviations of the 100x100 squares into which the original data were divided 
(see \S \ref{problem}); the simulation {\it does not} mimic the large scale 
structure of the data that manifests itself in such spreads.
Table 3 lists the mean and 
pixel-to-pixel rms of the ``mother'' M~31--Inner simulation; a theoretical 
``sky background'' produced by the WFPC2 Exposure Time Calculator 
\citep[ETC;][]{bire01} is included for easy comparison with the data.

When changing resolution, two sub--types of simulations were implemented. In one case, we employed
fully sampled Gaussians; in the other, we used Tiny Tim \citep{kris01} to 
produce a PSF, again appropriate for a G5V star in the F555W passband, again the
same for all stars. 
For the second sub--type, depending on the simulated 
resolution, we used either the model PSF 
for the center of the ACS-WFC, or that for the center of the 
ACS-HR camera. 
We produced once more only one square per angular resolution step. 
For simulations with Gaussian PSFs, we always produced
100x100 squares; for simulations with realistic PSFs,  
the sides of the squares scaled as the
inverse of the pixel sizes. In all cases, we used the ``tiling'' technique to 
make frames 80 \arcsec on the side.
The final step of the simulations was to add the HDF-N chips and 
count how many galaxies could be seen, on average, in these simulated images the 
size of the WFC field of the WFPC2 (5.3 arcmin$^2$).       

\subsection{LMC} \label{lmcsimu}

As a basis for the LMC simulations, we used the luminosity function published 
by \citet{jahr97} for nearby stars, but corrected as in \citet{holt97} for the 
LMC; that is, numbers of stars with $M_V$ = 7 or fainter have been multiplied 
by 1.6. We assumed a distance modulus to the LMC of (M-m)$_o$ = 18.6 
\citep{groe00}, and an extinction correction of A$_V$ = 0.63 \citep{elso97}. 
The luminosity function $\Phi (M_V)$ is presented in Table 4, in the same 
units used by \citet{jahr97}, or stars within 20 pc in 1 mag bin.

In the case of these simulations, we did not try to match the characteristics 
of the original data, but those of WFPC2, chip 3, {\it after} editing out the 
saturated stars. These are labeled WF3-E in Table 2. There are three other 
differences with respect to the M~31--Inner simulations. First, we used a Tiny 
Tim PSF appropriate for an F6V star through the F606W 
filter for the simulations with a ``realistic'' 
PSFs. Second, the number of total stars in a square 80\arcsec\ on the side is 
such ($<$ 2x10$^4$) that we had enough memory to generate a field in one pass
of {\it starlist}, as opposed to having to cover it with many copies of the 
same ``tile.'' This led to the third difference, i.e., that we 
attempted to match with the ``mother'' simulation not only the 
mean surface brightness and the mean standard deviation of the 
data, but also the spread of surface brightnesses. 
The ``mother'' LMC simulation has 0.027 stars pixel$^{-1}$
(or $\sim$ one star every 6 linear pixels, very close to our back of the
envelope calculation in \S \ref{motiv}),
$\mu_V$ = 21.68, FWHM = 0.29 mag (26\% of the mean), and 
$\sigma$ = 4.79 mag (441\% of the mean\footnote{This is 20 \% too high compared to the edited data,
but we could not improve it and keep the chosen luminosity function at the same time.}).

\subsection{Virgo} \label{virgosimu}

Here, we simulated four different situations based on the NGC 4536 data: an 
interarm region with the inclination to the line of sight of NGC 4536 
(65\degr); an arm, also with such inclination, a face--on interarm region, and 
an arm in a face--on galaxy. For the interarm regions, we used the same 
stellar mixture as for the disk of M~31; for the face--on simulations, 
we ``deprojected'' the $i$=65\degr\ populations as if they were
optically thin.  Interestingly, given that the 
inclination of NGC~4536 is smaller than that of M~31 (78\degr), we needed the 
equivalent of 60 stars (instead of 50 or fewer) at the distance of Andromeda 
to match the characteristics of the NGC~4536 data. We assumed an {\it 
apparent} distance modulus to Virgo of (m-M)$_{A_V}$ = 31.23 \citep{saha96a}; 
given the {\it apparent} distance modulus to M~31 of (m-M)$_{A_V}$ = 24.82 
(see \S \ref{m31simu}), we were trying to simulate a very similar stellar 
population located $\sim$ 19 times further away. Naturally, computer memory 
limited us to 5x5 pixel ``tiles'' for the ``mother'' simulation of the 
``inclined'' Virgo galaxy; however, these were so small that, at each resolution
step, we produced 400 of these ``sub--tiles'' to generate a 100x100 pixel 
``tile.'' We used the same size of ``tile'' for the face--on 
cases, even though (assuming the same population) these had 2.5 times fewer 
stars per unit area [cos(65\degr) $\sim$ 0.42].

In order to mimic ``arms,'' we kept the old MS and the red giants 
in the same numbers as for the ``interarm'' simulation, but this time we added 
a young main sequence with an upper type B3V (10 M$_\odot$), and $\sim$ 2.4 \% 
by mass of the old main sequence. For stars with masses between 9 M$_\odot$ 
and 10 M$_\odot$, this is approximately equivalent to a density of 0.01 stars 
pc$^{-2}$. \citet[and references therein]{hunt95} have found densities between 
0.007 and 0.02 massive stars pc$^{-2}$ for OB associations and selected 
regions of the Milky Way, the LMC, I Zw18 and M~31. Interestingly, we did not 
need to increase the numbers of the old population stars, as one would in 
principle have to do in the presence of a density wave; pushing up the 
fraction of old population augmented the mean surface brightness while 
decreasing the standard deviation under that shown by the data. As for M~31--Inner, the 
parameters and characteristics of the ``mother'' simulations are given in 
Table 3.

\subsection{M~31 radial simulations} \label{m31_nsimu}  

As we mentioned earlier, we have simulated the changes in surface brightness 
with radius of the disk of M~31, at the resolution and with the PSF of the 
WFPC2 on board {\sl HST}. We have used the stellar mixture that worked for the
M~31--Inner field, and changed the number of stars per pixel to reproduce the 
different surface brightnesses found between 10$'$ and 100$'$ galactocentric \
radius, at 10$'$ steps. The surface brightness profile has been taken from 
\citet{walt88}. The radial run of surface brightnesses numbers of old MS stars per
pixel are given in 
Table 3.

\section{Results} \label{results}

Figures \ref{m31infig}--\ref{ronfig} show the results of our simulations; figures \ref{m31infig}--\ref{virgosim} present 
information in basically the same fashion. Background galaxy counts observed 
are plotted vs.\ number of stars per pixel. The field of view of 5.3 arcmin$^2$
is that of the combined 3 WFC subimages of the WFPC2. The left panels display the 
number--counts obtained with ``realistic'' Tiny Tim PSFs, while the right 
panels show results for Gaussian, fully--sampled PSFs (except for the leftmost 
square points, which show once more results for WFPC2 aboard {\sl HST}). Empty 
symbols are for ``noiseless'' (infinite integration time) data, whereas full 
symbols are for 7200 s (figures \ref{m31infig} and \ref{virgosim}, for M~31--Inner and Virgo, 
respectively) or 1800 s (Figure \ref{lmcfig}, LMC) exposures with {\sl HST}; these times 
should be scaled by (2.4/D)$^2$ for telescopes with a different aperture D in 
meters. It is indicated when results correspond to diffraction limited 
telescopes. For the ``noisy'' data, we have used the gain and read--out noise 
of the WFPC2 camera, {\it except} for the ACS simulations in the left panels, 
where values anticipated for the ACS were applied \citep{pavl01};
the nominal, foreseen gain has been multiplied by 3, given that the ACS
is expected to be 3 times faster than the WFPC2. In all cases,
sky backgrounds were taken from the WFPC2 ETC \citep{bire01}.   
For reference, in the absence of any foreground object and using --only in this case-- 
the photometry package SExtractor \citep{bert96}, we have identified 
$\sim$ 1700 galaxies, with an isophotal $V$ detection limit of
27 mag$_{AB}$ arcsec$^{-2}$, in the 3 WFC chips of the virtually noiseless HDF-N. 

In the past \citep*[Paper I;][]{gonz99}, we have derived total (Poisson plus 
clustering) errors for the galaxy counts, as a function of total magnitude 
completeness limit, from values of the two--point correlation function, 
$\omega(\theta)$, published in the literature \citep*[Roche et al.\ 
1993;][]{brai95}. These theoretical total errors amount to $\sim$ 
2$\times\sigma_{Poisson}$; $\sigma_{Poisson} = N^{1/2}$,where $N$ stands 
for the galaxy counts. However, in \citet{gonz99} we found empirical errors 
reaching close to twice this value. Interestingly, \citet[columns 3 and 4 of 
their Table 2]{roch93} obtain similar empirical errors, albeit with 
significantly larger fields ($\sim$ 23.5 arcmin$^2$). Their predicted error 
ranges between 1.5--2$\times\sigma_{Poisson}$, while their empirical error 
varies between 2.5--3.5$\times\sigma_{Poisson}$. Tracking down the reason for 
this discrepancy is beyond the goals of the present work but, in the meantime, 
we will adopt $\pm$3.5$\times\sigma_{Poisson}$ error bars. This value should 
be conservative enough, as well as convenient, since in all the simulations we have identified background 
galaxies by eye (\S \ref{problem}) and hence only could have an approximate estimate of 
the detection completeness limits.

Simulations for the M~31-Inner field with realistic PSFs (Figure \ref{m31infig}a) and exposure times
show that, while the ACS will multiply the number of background galaxies that
can be detected by two compared to WFPC2, then there will be no significant 
further improvement until we can count with a diffraction--limited 10--m 
optical telescope. The galaxy counts in the simulations with the Gaussian PSFs 
(Figure \ref{m31infig}b), however, improve steadily with the size of the telescope 
(although here we invite the reader to notice the jump in the counts around 
1 star per pixel); already 
a diffraction limited 3--m telescope would be almost as good as the 10--m 
telescope in the left panel, or 5 times better than WFPC2. The 4.5--m 
telescope with a Gaussian PSF would be 10 times better than today's WFPC2. 
With our extremely conservative error bars, we would expect the uncertainty 
on the extinction determination to decrease from $\sim\pm$ 1.2 mag with WFPC2
to $\pm$ 0.9, $\pm$ 0.5, and $\pm$ 0.4 mag, respectively, with 2, 5, and 10 
times more galaxies.

The LMC represents the most dramatic case. Figure \ref{lmcfig}a shows basically constant 
number--counts; the ACS will already be close to the meager factor of two 
improvement we expect from a diffraction--limited in the optical, 10--m telescope.
Figure \ref{lmcfig}b again shows a steady increase in the number--counts with resolution, 
with a factor of 8 more counts already expected for a 3--m diffraction limited 
telescope! It also illustrates the situation where one 
benefits the most from long exposure times. The values above translate into uncertainties in the extinction of
$\sim\pm$ 0.8 mag now, $\pm$ 0.5 mag with the ACS, and $\pm$ 0.3 mag with 8 
times more background galaxies. 

Figure \ref{virgosim} displays results for a Virgo disk. The four top panels (a--d) 
summarize simulations for a galaxy with an inclination to the line of sight 
of 65\degr; the bottom panels (e--h) correspond to a face--on disk. Finally, 
panels (a--b) and (e--f) are for ``interarm'' regions, while panels (c--d) 
and (g--h) represent ``arms.'' The most interesting lesson from this figure is 
that, when using ``realistic'' exposure times, all panels are basically flat, 
with improvements that are, at most, of a factor of two. Also, for each 
different case (from top to bottom: interarm inclined, arm inclined, interarm 
face--on, arm face--on), the results are the same regardless of the PSF used. 
Thus, in the case of Virgo we are already close to the best results we will be 
able to get with the present day {\sl HST}. The extinction uncertainties range 
in most cases between $\pm$ 0.5 and $\pm$ 0.3 mag, in perfect agreement with 
Paper I.  

Finally, Figure \ref{ronfig} shows background galaxy counts that should be 
observable at different radii through the disk of M~31 with WFPC2 data, 
assuming everywhere the same population mixture used to simulate the 
M~31---Inner field. Numbers of 
stars per pixel and $\mu_{V}$ from \citet{walt88} 
are also shown. All the simulations in this series employ a ``realistic'' 
Tiny Tim PSF; ``noisy'' ones (empty symbols) have exposure times of 1800 s.   
With existing data, it will be hard to reduce the extinction uncertainty 
below $\pm$ 0.5 mag, even with long exposure times and for the largest radii 
investigated. 

Table 5 lists, for each simulation, the pixel size, the linear size of 
each ``tile,'' the number of stars per pixel, the PSF used, 
the background added, exposure time, 
gain, read--out noise, and galaxy counts in a field of 5.3 arcmin$^2$. 
For comparison, the number--counts in a 
simulation where the HDF has been added to the real data are also shown.

\section{Discussion} \label{disc}

As hypothesized initially, the granularity of the foreground galaxy 
(as measured by the number of stars per pixel) does appear to play a key role 
in the detectability of background galaxies. However, the comparison between 
diffraction--limited telescopes with realistic and Gaussian PSFs (Figures 
\ref{m31infig}--\ref{virgosim}) shows that the ``quality'' of the PSF often plays a 
significant role as well. This points to two distinct sources of granularity 
in the images. A ``primary'' granularity due to the stellar distribution 
itself -- and best measured in the simulated images with Gaussian PSFs -- and 
a  ``secondary'' granularity produced by the wings and spikes of the PSF. 
At the distance of Virgo, the 
number of stars per pixel is always very large or large (between 50 and 10$^4$,
depending on resolution), and each individual star is very faint, so the 
stellar content of the foreground galaxy appears as a smooth continuum with 
very little sub--structure (i.e.\ almost without primary granularity). Changing the 
quality of the PSF (from a perfect Gaussian to a realistic case) has almost
no influence, because the additional ``secondary'' granularity introduced
by an imperfect PSF has essentially as little sub--structure as the stellar 
field itself. 

The situation in the LMC is almost exactly opposite to that of Virgo. In the LMC,
there is only one star every several tens or hundreds of pixels
(again, depending on resolution); in addition, owing to the 
proximity of the LMC, this star is often bright. Although this 
implies a ``primary'' granularity significantly larger than in Virgo, one 
would {\it a priori} expect to be able to see ``between'' the stars. Only a 
few pixels out of several tens or hundreds would image (and, therefore, be 
affected by) stars proper in the LMC, and most pixels would be expected to be free 
of such emission. That so few background galaxies can be identified in the 
LMC implies that, there, the secondary granularity dominates completely 
over the primary one. A quite natural explanation to this phenomenon is that a realistic 
(non-Gaussian) PSF will spread a significant fraction of the stellar light 
over many pixels, in effect ``propagating'' the granularity 
to pixels that would otherwise be emission--free. For a 10-m 
diffraction--limited telescope, up to 90\% of the background galaxies can be 
lost to this secondary granularity (Figure \ref{lmcfig}). 
Figure \ref{lmc_acs} demonstrates this dramatic effect for the case 
of the ACS-HR camera (i.e., a diffraction limited 2.4-m telescope). 
This should be a good general example because the PSF of this instrument
is well sampled (which is also the reason why we have used the 
ACS-HR PSF to simulate images obtained with larger telescopes). 
For reference, the top panel of this figure shows WF3 subfield of the 
WFPC2 F606W data of the LMC. The middle left panel presents our simulation
of the same field with the resolution of the ACS-HR and the Tiny Tim 
PSF for said camera and wavelength (we remind the reader that we 
have omitted saturated stars 
from the simulation). The middle right panel shows our simulated field with 
a realistic PSF plus the HDF-N; like in the case of the 
right panel of figure \ref{lmc_color}, 
reference galaxies can only barely be seen. The bottom panels display,
respectively, the simulation with a Gaussian PSF without ({\it left}) and with 
({\it right}) the HDF-N added in; tens of background reference galaxies can easily
be counted in the right bottom panel. 
Figure \ref{psfs} shows linear ({\it a} and {\it c})
and logarithmic ({\it b} and {\it d}) surface plots of the ACS-HR ({\it top})
and Gaussian ({\it bottom}) PSFs. As can be seen in the linear plots,
the cores are similarly sampled, but      
the core of the Gaussian out to 3$\sigma$ receives 97\% of the total light,
whereas only 58\% enters in the same area of the ACS-HR PSF.
The wings of the PSFs are better appreciated in the logarithmic plots;
the light in the wings and spikes of the realistic PSF
reduces the signal--to--noise ratio of background galaxies and explains
the enormous disparity in the efficiency with which they are identified
in both cases.  A bad subtraction of the realistic wings can only 
introduce more noise and further reduce the detectability of background objects.

The case of M~31 is intermediate between Virgo and the LMC. There, 
changing the resolution of the telescope helps little, until the number
of stars per pixel falls below about 1 where a jump can clearly be seen on 
Figure \ref{m31infig}. This jump corresponds to a change of regime: with less than 
1 star per pixel, one encounters a situation similar to that of the LMC, 
where the secondary granularity becomes strongly dominant. Indeed, the 
difference between Gaussian and realistic PSFs becomes important only in that
regime. With a few bright stars per pixel, the primary granularity is 
the dominant factor. This is also quite natural, because 
this is precisely the case where all pixels image stars, and at the same 
time large
pixel-to-pixel variations
(primary granularity) are expected. This situation, where each pixel contains
more than 1 star, and where stars cannot be identified independently or their
positions within a pixel determined accurately, 
essentially corresponds to the regime of surface brightness fluctuations 
\citep[SBFs;][]{tonr88}. 

The comparison between M~31
and Virgo shows that SBFs largely prevent the detection of background 
objects only in Local Group galaxies, where each pixel contains  
a small number of bright stars.  
At the same time, the simulations with 
realistic PSFs show that there is always at least a 
slight improvement in the detectability of background galaxies with  
better resolution (which appears paradoxical, since fewer 
stars per pixel seem to imply more granularity). 
This improvement, moreover, is significantly smaller in Virgo 
than in the LMC. All these facts can be understood in a statistical way.
For simplicity, we will 
assume that all stars are identical, and have an intrinsic flux $f_*$ (in 
photons sec$^{-1}$ received at the detector). The total number of photons 
$\gamma$ received in each pixel of the CCD in an integration time $t$, assuming that 
there are $n$ stars per pixel is:

\begin{equation}
\gamma~=~ntf_*
\end{equation}

\noindent
The integration time $t$ can be considered a constant here, but both $f_*$
and $n$ contribute to the noise. Therefore, neglecting the sky background 
and the read-out noise (RON) of the detector, the total noise is:

\begin{equation}
\sigma~=~\sqrt{\left(\sigma_nf_*t\right)^2+\left(n\sigma_{f_*}t\right)^2}
\end{equation}

\noindent
Since the uncertainty on $f_*$ is $\sqrt{f_*}$, and that of $n$ is $\sqrt{n}$, we
obtain the following expression for the total noise:

\begin{equation}
\sigma~=~\sqrt{nf_*^2t^2+n^2f_*t^2}~=~t\sqrt{nf_*^2+n^2f_*}
\end{equation}

\noindent
A background galaxy will be detected with a
signal-to-noise ratio equal to the  
quotient of the number of photons received from that background object 
($f_{bg}t$) divided by the noise computed above (the Poisson noise due to the background 
galaxy can be neglected, since the surface brightness of such objects is always so much fainter than 
the foreground stars):

\begin{equation}
\left(\frac{S}{N}\right)_o~=~\frac{f_{bg}t}{t\sqrt{nf_*^2+n^2f_*}}~=~\frac{f_{bg}}{\sqrt{nf_*^2+n^2f_*}}
\end{equation}

\noindent
When the pixels are made $L$ times smaller (on the side), the flux received 
from the background galaxy ($f_{bg}$) and the number of stars per pixel $n$
both diminish by $L^2$ (the background galaxy is assumed to be resolved -- if 
it was not, it could not be identified as such in our simulations), but the 
flux of each star stays constant (given that stars are unresolved and always
fall completely in one pixel). The
signal-to-noise ratio {\it per pixel}, therefore, becomes:

\begin{equation}
\frac{S}{N}~=~\frac{f_{bg}}{\sqrt{L^2nf_*^2+n^2f_*}}
\end{equation}

\noindent
But the pixels are now $L$ times smaller on the side, so the background 
galaxy now occupies $L^2$ more pixels; and since the noise adds in  quadrature, 
the background galaxy is now detected with a total signal--to--noise ratio: 

\begin{equation}
\left(\frac{S}{N}\right)_L~=~\frac{Lf_{bg}}{\sqrt{L^2nf_*^2+n^2f_*}}
\end{equation}

\noindent 
Finally, if the same foreground stellar population is pushed $d$ times farther away,
the number of stars per pixel $n$ is multiplied by $d^2$, while the flux of 
each star $f_*$ is divided by $d^2$. The flux of the background galaxy 
obviously remains unchanged, and the signal-to-noise ratio becomes:

\begin{equation}
\left(\frac{S}{N}\right)_d~=~\frac{Lf_{bg}}{\sqrt{\frac{L^2nf_*^2}{d^2}+n^2f_*d^2}}
\end{equation}

The behavior of this expression as a function of pixel size (set by $L$) for 
the same stellar population at three different distances $d$ is shown on the 
left panel of Figure \ref{eq7}. The value of the various parameters have been chosen to 
reproduce (although with arbitrary units) the conditions found in the LMC, M~31 
and Virgo. The exact normalization factors will depend on the way 
signal-to-noise and background galaxy surface brightness detection limits are related, 
an issue that we do not investigate here. Figure 
\ref{eq7}b shows again the number of galaxies that are detected in the LMC, M~31, and Virgo as 
a function of the pixel size (taken from Figures \ref{m31infig}--\ref{virgosim}). Those numbers are the 
results of the simulations with infinite integration time, and Gaussian PSFs 
-- the best comparison data, since integration time, 
sky background, RON, and PSF 
``imperfections'' are 
not accounted for in Equations 1--7. 
Also, an arbitrary constant has been added to the logarithm of the    
number of background galaxies seen through each object, to    
facilitate the comparison of the data with the model. The qualitative agreement between the left and right panels 
of Figure \ref{eq7} is 
remarkable, especially given the extremely simple model we 
have adopted for the noise; 
in addition to illustrating the origin of the detection efficiency 
of background galaxies, the concordance between model and observations 
validates the approach used to count the galaxies.
The number of background galaxies detected behind the LMC increases 
rapidly as the pixels are made smaller, until it reaches a plateau. This 
plateau corresponds to the situation where the signal--to--noise ratio of the
galaxy detection is dominated by ``the space between the stars,'' or the 
smooth sky background in real data; the number of 
background objects seen 
is not improved with better resolution, since 
all the galaxies detectable with the chosen integration time 
are already identified. 
In M~31, the detectability also increases rapidly (although less 
rapidly than in the LMC), but never reaches a plateau (note that it would, 
eventually, if 
the pixels were made even smaller). In Virgo, the improvement in detectability
as pixels are made smaller still occurs, but is much less pronounced than in 
the LMC or M~31, precisely because the foreground is approximately 
smooth. The two main features of the detectability of background 
galaxies in the simulations (increase followed by leveling--off in the case of 
the LMC; and smaller slope of the improvement for farther objects) are, thus, reproduced by 
Equation 7. Our approach of considering separately the contributions of 
$n$, the number of stars, and $f_*$, the flux of each star, to the noise,
also allows us to explain away the paradox stated at the beginning of the
section. For an individual foreground object, the increasing number of
stars per pixel as resolution worsens translates into smaller pixel--to--pixel
fluctuations, which ought to lead to an {\em increase} in the counts of 
background galaxies. But of course (even though the surface brightnesses of
both foreground and background galaxies stay constant), the total brightness of each pixel is
also increasing, which means an increase in photon noise per pixel and a reduction
in the background galaxy contrast, leading to a {\em reduction} in the counts
of identified background galaxies. Of the two competing effects, the
simulations show that the latter ``wins.''  

In the case of Virgo, we are, and will remain for the 
foreseeable future, within the regime of faint SBFs.
Better resolution will yield only small improvement,
and results will be independent of the quality of the
PSF (Fig.\ \ref{virgosim} and Fig.\ \ref{eq7}).
Conversely, in the LMC, we are already outside of the (bright) SBF 
regime with $HST$ WFPC2 images; 
in theory, improvements in resolution 
should translate into enormous gains in galaxy detection
(Fig.\ \ref{lmcfig}b and Fig.\ \ref{eq7}),
but Fig.\ \ref{lmcfig}a shows that these will not be obtained with
a realistic PSF. The successful application of the SFM in this
case will require an adequate subtraction of the foreground stars 
{\em with their diffraction pattern}, a 
notoriously difficult task. 
In the case of WFPC2 data, the difficulty arises from the 
undersampling, 
chromaticity, and spatial 
and time variability 
of the PSF. 
Model PSFs, such as those produced by Tiny Tim, are clearly not sufficiently 
accurate, especially in the wings. Better, 
possibly iterative, procedures will have to be developed
(\citealp{ande00}, and references therein). 
An attempt to tackle WFPC2 images of the LMC 
in this way will be presented in 
a subsequent paper \citep{gonz03}. It should be emphasized,
however, 
that much work will probably be needed in order to fully master 
optical/near-IR PSF subtraction
procedures, both for WFPC2 and for future telescopes and instruments. 
For ACS data, for instance, aside from the undersampling 
of the PSF by the WFC, the biggest problem we can anticipate is
the huge geometric distortion of the field--of--view. 

As for M~31, Fig.\ \ref{m31infig} and Fig.\ \ref{eq7} show that small improvements
can still be obtained with better resolution, regardless of
PSF quality. Furthermore, 
images obtained with a 10-meter 
diffraction-limited optical telescope (which 
could be available in the not--so--distant future) will 
have less than 1 star per pixel.
Except for the apparent brightness of the stars, such images of M~31 will be 
similar to the existing $HST$ images of the LMC, where the detectability 
of background galaxies is limited by the secondary granularity produced by 
structure in the PSF. Accurate removal of individual
stars will be required at that point to improve the
results of the SFM. For existing WFPC2 data, one could propose 
to remove at least the brightest stars. It would be very difficult,
if not impossible, to do a good job, since this would require the accurate
determination of both the position of the stars and the shape of 
the PSF in a very crowded field, with many more than 1 star per pixel.  
To investigate whether it would be worthwhile to pursue this idea, 
we generated an artificial M~31--Inner 
image with a list of stars from which we deleted the red giants and 
young MS stars brighter than magnitude 25. The identified background galaxies in 
a simulated 5.3 arcmin$^2$ field went up from 18$\pm$15 (Table \ref{tsim2}) to 
26$\pm$18; the uncertainty in the extinction would go  
down from 1.3 to 1.1 mag. 
One could not remove fainter stars, since incompleteness sets in 
at 25 mag (Fig.\ \ref{cmfig}).
 
Two final comments should be made here. First,
integration times are {\em never} an issue for realistic PSFs. The results 
of the simulations for ``short'' (0.5--2 hours) integration times are always 
nearly as good as those with infinite integrations. 
Second, 
a note of caution. We have seen here that, 
using {\sl HST} WFPC2 data, 
background galaxies can be detected through the disks of Virgo galaxies, but 
not through Local Group galaxies.
It could, therefore, be proposed to simply smooth the data on Local Group 
galaxies to match the resolution achieved by {\sl HST} in Virgo. This -- of
course -- would be foolish, because it would also smooth the background 
galaxies. What one needs is a smooth foreground stellar field, but sharply 
defined background objects. 

\section{Conclusions}

The analysis of the simulations presented in this paper shows that, for the time being,
Virgo seems to offer the best prospects for the application of the SFM. Reasonably accurate 
statements can be made about the opacity of the disks at that distance (Paper 
I), and long integration times are not required. Arcsecond resolution \hi\ 
VLA observations could be obtained for a significant number of the spirals 
in Virgo, and the Atacama Large Millimeter Array (ALMA) will be able to 
provide high--sensitivity CO maps with the same resolution. Obtaining the 
opacity of a large sample of spiral in Virgo using the SFM could, therefore, 
be used to statistically investigate the properties of the ISM there. 
Objects more 
distant than Virgo would tend to present 
even smoother foreground stellar fields, 
but the opacity there would be determined with less accuracy because their 
smaller angular size would reduce the number of background galaxies that could 
be identified.

For Local Group galaxies like M~31 and the LMC, the SFM
will hardly be usable, even with telescopes 
significantly better than those currently at hand,
unless the PSF is better understood. 
An adequate subtraction of individual point sources from 
WFPC2 data
is an extremely difficult task, owing both to the undersampling \citep{ande00}, and to the
variability of the PSF with time, position in the chip, and spectral energy
distribution of the imaged objects; the variability affects especially our 
present 
ability to model the wings \citep{kris01}, i.e., 
the source of the secondary granularity
that hampers the detection of the background galaxies. 
Since the quality and stability of the PSFs of future instruments
are unlikely to be much better than those of the WFPC2
PSF, a detailed 
understanding of optical/near-IR PSFs will be required to 
improve the results of the SFM in any Local Group galaxy.
 
\acknowledgments
R.A.G. thanks M.\ Gu\'elin and the Institut de Radio-Astronomie 
Millim\'etrique (IRAM), Grenoble, for their hospitality during the 
preparation of this work; we are grateful to B.\ Turner and T.\ Dame for 
locating and faxing the Shapley 1951 reference within minutes of our 
request. R.A.G.\ was partially funded by the following grants: 
STScI AR-08360.01-A, NSF INT-9902653, and Mexican 
CONACyT I-35675E and 593-2000. L.L.\ acknowledges the support of 
CONACyT grant 592-2000.

\clearpage

\begin{figure}
\hspace*{2.5cm}
\caption{{\it (a), (c), and (e): cutouts of the WF3 frame of the LMC.} 
The color mosaics, 
12\farcs8 on the side, were 
produced by combining together the F606W frame, the F814 frame, and
an average of the two; they are displayed in a linear scale. {\it (b),
(d), and (f): 
same as left panels, with the WF3 frame of the HDF-N added in.} It is extremely
hard to distinguish the simulated reference galaxies; one of the very
brightest galaxies in the HDF-N can be seen in panel (f).}
\label{lmc_color}
\end{figure}

\begin{figure}
\caption{Comparison between 1-D images with 
negligible large-scale structure, but large pixel-to-pixel variations
(right panels), and 1-D images with important
large-scale structure, but small pixel-to-pixel variations
(left panels). The pair of numbers given at the top-left of
each sub-panel represents the mean/rms in the
considered section of the image. All units are arbitrary.
While the mean and rms remain unaffected by the size of the
portion of the image considered in the case with no 
large-scale structure (right panels), they change significantly 
when large-scale structure is present (left panels).}
\label{rmsfig}
\end{figure}

\begin{figure}
\caption{V vs.\ V-I color-magnitude diagram for stars in the 3 wide-field 
chips of the WFPC2 M~31--``Inner'' image. The box delineates area of CMD 
populated by B stars. B5V stars have $m_V$ = 23.7 in this diagram.}
\label{cmfig}
\end{figure}

\begin{figure}
\caption{Galaxy counts in the M~31--Inner field simulations. The field of 
view of 5.3 arcmin$^2$ is equal to that of the combined 3 wide--field chips 
of the WFPC2. {\it (a):} ``realistic'' Tiny Tim PSF. {\it (b):} Gaussian, 
fully--sampled PSF (except for WFPC2 simulation). {\it Empty symbols:}
noiseless data. {\it Full symbols:} 2--hour exposures with {\sl HST};    
this time should be scaled by (2.4/D)$^2$ for telescopes with a different 
aperture D in meters. Error bars are $\pm$3.5$\times \sigma_{Poisson}$ (see text).}
\label{m31infig}
\end{figure}

\begin{figure}
\caption{Galaxy counts in the LMC field simulations. The field of view of 
5.3 arcmin$^2$ is equal to that of the combined 3 wide--field chips of the 
WFPC2. {\it (a):} ``realistic'' Tiny Tim PSF. {\it (b):} Gaussian, 
fully--sampled PSF (except for WFPC2 simulation). {\it Empty symbols:}
noiseless data. {\it Full symbols:} 1800 s exposures with {\sl HST};
this time should be scaled by (2.4/D)$^2$ for telescopes with a different 
aperture D in meters. Error bars are $\pm$3.5$\times \sigma_{Poisson}$.}
\label{lmcfig}
\end{figure}

\begin{figure}
\hspace*{1.5cm}
\caption{Galaxy counts in Virgo galaxy simulations. The field of view of 
5.3 arcmin$^2$ is equal to that of the combined 3 wide--field chips of the 
WFPC2. {\it Left panels:} ``Realistic'' Tiny Tim PSF. {\it Right panels:} 
Gaussian, fully--sampled PSF (except for solid squares). {\it Empty symbols:}
noiseless data. {\it Full symbols:} 2--hour exposures with {\sl HST};    
this time should be scaled by (2.4/D)$^2$ for telescopes with a different 
aperture D in meters. {\it (a) and (b):} ``interarm'' region, inclined 65\degr 
to the line of sight. {\it (c) and (d):} ``arm,'' inclined 65\degr to the 
line of sight. {\it (e) and (f):} face--on ``interarm'' region. {\it (g) and 
(h):} face--on ``arm.'' Error bars are $\pm$3.5$\times \sigma_{Poisson}$.}
\label{virgosim}
\end{figure}

\begin{figure}
\caption{Galaxy counts in WFPC2 simulations of M~31 stellar foreground at 
different galactic radii. Stars pixel$^{-1}$ and V surface brightness are 
also indicated. {\it Empty symbols:} noiseless data. {\it Full symbols:} 
1800 s exposures. Error bars are $\pm$3.5$\times \sigma_{Poisson}$.}
\label{ronfig}
\end{figure}

\begin{figure}
\hspace*{2.5cm}
\caption{Simulations of the LMC with Gaussian and realistic PSFs.
{\it (a):} WF3 subfield of the
WFPC2 F606W data of the LMC. {\it (b):} simulation
of the same field with the resolution of the ACS-HR and the Tiny Tim
PSF for said camera and wavelength (we
have omitted saturated stars
from the simulation). {\it (c):} simulated field with
a realistic PSF plus the HDF-N; 
reference galaxies can only barely be seen. 
{\it (d):} simulation with a Gaussian PSF.
{\it (e):} simulated field with a Gaussian PSF and 
the HDF-N added in; tens of background reference galaxies can easily
be counted in this panel. All images are 75\arcsec on the side; 
none of the simulations has Poisson noise.}
\label{lmc_acs}
\end{figure}

\begin{figure}
\hspace*{-1cm}
\caption{Surface plots of 
PSFs. {\it (a)}: linear plot of the ACS-HR Tiny Tim PSF; {\it (b):}
logarithmic plot of the ACS-HR Tiny Tim PSF;
{\it (c):} linear plot of fully sampled Gaussian PSF for the
same resolution; {\it (d):} logarithmic plot of fully sampled  
Gaussian.} 
\label{psfs}
\end{figure}

\begin{figure}
\caption{Comparison of the behaviour of Equation 7 ({\it left panel}) with the 
number of background galaxies detected
through the LMC, M~31, and Virgo ({\it right panel});
an arbitrary constant has been added to the logarithm of the 
number of background galaxies seen through each object, to 
facilitate the comparison of the data with the model.}
\label{eq7}
\end{figure}

\newpage

\include{Gonzalez.tab1}

\include{Gonzalez.tab2}

\include{Gonzalez.tab3}

\include{Gonzalez.tab4}

\include{Gonzalez.tab5}

\include{Gonzalez.tab6}

%
%
%
%
%
%
%
%
%
%
%

\end{document}

%% file: Gonzalez.tab1.tex
\begin{deluxetable}{lccrrr}
\tablewidth{5.60in}

\tablecaption{Observation Log}
\tablehead{
 & & & \multicolumn{3}{c}{Exposure\ time (s)}
\\[-0.3cm]
 &$\alpha$ &$\delta$ & \multicolumn{3}{c}{\hrulefill} \\%
\colhead{Field (archive descriptor)} &
\colhead{(J2000)} &
\colhead{(J2000)} &
\colhead{F555W} &
\colhead{F606W} &
\colhead{F814W}\\[-0.3cm]
}
\startdata

LMC (PAR)\dotfill & 05\ 35\ 38.5& -69\ 24\ 18& &1000\ \ \ \ &1000\ \ \ \ \\
M~31-D478 (PAR)\dotfill & 00\ 42\ 52.1&\ 41\ 24\ 53& &5600\ \ \ \ &6400\ \ \ \ \\
M~31-Inner\dotfill & 00\ 44\ 23.7&\ 41\ 45\ 16&2400\ \ \ \ &&2500\ \ \ \ \\
NGC~4536\dotfill& 12\ 34\ 27.1&\ 02\ 11\ 14& 68000\ \ \ \ && 20000\ \ \ \ \\
HDF\dotfill & 12\ 36\ 49.4 &\ 62\ 12\ 58 & & 109050\ \ \ \  & 123600\ \ \ \ \\

\enddata
\tablecomments{Units of right ascension are hours, minutes,
and seconds; units of declination are degrees, arcminutes, and arcseconds.
When different from our adopted field name, the {\sl HST} archive
descriptor is given in parentheses.}
\label{tobs}
\end{deluxetable}

%% file: Gonzalez.tab2.tex
\begin{deluxetable}{lclcllr}
\tablewidth{5.6in}
\tablecaption{Characteristics of Observed Subfields}
\tablehead{
\colhead{Field} &
\colhead{Band} &
\colhead{Subfield} &
\colhead{$\mu_V$} &
\colhead{FWHM} &
\colhead{$\sigma$} &
\colhead{ZP}\\[-0.3cm]
}
\startdata

NGC 4536\dotfill &F555W&WF2 (arm)& 21.35 &0.27\ (25\%)&0.08\ (\hspace*{0.38cm}7\%)&22.57 \\
&&WF3 (arm)& 21.33 &0.27\ (25\%)&0.08\ (\hspace*{0.38cm}7\%)&22.55 \\
&&WF4 (interarm)& 21.38 &0.12\ (12\%)&0.04\ (\hspace*{0.38cm}4\%)&22.52 \\
M~31-D478\dotfill &F606W&WF2&20.70&0.10\ (\hspace*{0.19cm}9\%)&0.39\ (\hspace*{0.19cm}36\%)&23.11 \\
&&WF3&20.39&0.18\ (17\%)&0.35\ (\hspace*{0.19cm}32\%)&23.11 \\
&&WF4&20.25&0.15\ (14\%)&0.32\ (\hspace*{0.19cm}30\%)&23.09 \\
M~31-Inner\dotfill &F555W&WF2&21.52&0.00\ (\hspace*{0.19cm}0\%)&0.36\ (\hspace*{0.19cm}33\%)&22.57 \\
&&WF3&21.50&0.00\ (\hspace*{0.19cm}0\%)&0.32\ (\hspace*{0.19cm}29\%)&22.55 \\
&&WF4&21.47&0.01\ (\hspace*{0.19cm}1\%)&0.36\ (\hspace*{0.19cm}33\%)&22.52 \\
LMC\dotfill &F606W&WF2&21.17&0.33\ (31\%)&4.67\ (430\%)&23.11 \\
&&WF3&21.53&0.25\ (23\%)&4.68\ (431\%)&23.11 \\
&&WF3-E&21.66&0.26\ (24\%)&3.98\ (369\%)&23.11 \\
&&WF4&21.28&0.36\ (33\%)&4.98\ (459\%)&23.09 \\

\tablecomments{Col.\ (4) mean of distribution of $V$ surface brightnesses  
(mag) of 100$^2$ pixel image sections. Col.\ (5) Full width at half
maximum (mag and percentage relative to mean)
of same distribution; diagnostic of large scale 
structure in subfield. Col.\ (6) mean of distribution
of pixel-to-pixel rms variations in 100$^2$ pixel 
image sections (mag and percentage relative to
mean surface brightness); diagnostic of image granularity/
telescope resolution. Col.\ (7) Zero point for conversion to 
$V$ magnitudes.}



\enddata
\label{tsq}
\end{deluxetable}

%% file: Gonzalez.tab3.tex
\begin{landscape}
\begin{deluxetable}{lrrrrr}
\tablewidth{7.85in}

\tablecaption{Parameters of the simulations for M~31--Inner and Virgo}
\tablehead{
 & M~31--Inner &  \multicolumn{4}{c}{Virgo}                               \\%
\vspace{-0.3cm}\\%
\cline{3-6}
\vspace{-0.3cm}\\%
 &             &  Iterarm ($i$ = 65\degr) & Interarm (Face--on)  &  Arm ($i$ = 65\degr)  & Arm (Face--on)  \\%
\vspace{-0.3cm}
}
\startdata 

Old stars per pixel\dotfill         & 50     \hspace*{1.0cm}     & 2.4 10$^4$ \hspace*{0.8cm} & 1.0 10$^4$ \hspace*{0.8cm} & 2.4 10$^4$ \hspace*{0.8cm} & 1.0 10$^4$  \hspace*{1.0cm}\\%
Old population M$_{up}$ \dotfill    & 3.3    \hspace*{1.0cm}     & 3.3        \hspace*{0.8cm} & 3.3        \hspace*{0.8cm} & 3.3        \hspace*{0.8cm} & 3.3         \hspace*{1.0cm}\\%
Old population M$_{low}$\dotfill    & 15.5   \hspace*{1.0cm}     & 15.5       \hspace*{0.8cm} & 15.5       \hspace*{0.8cm} & 15.5       \hspace*{0.8cm} & 15.5        \hspace*{1.0cm}\\%
Red giant stars per pixel\dotfill   & 0.025  \hspace*{1.0cm}     & 12         \hspace*{0.8cm} & 5          \hspace*{0.8cm} & 12         \hspace*{0.8cm} & 5           \hspace*{1.0cm}\\%
Red giant M$_{up}$\dotfill          & $-$0.8 \hspace*{1.0cm}     & $-$0.8     \hspace*{0.8cm} & $-$0.8     \hspace*{0.8cm} & $-$0.8     \hspace*{0.8cm} & $-$0.8      \hspace*{1.0cm}\\%
Red giant M$_{low}$\dotfill         & 0.6    \hspace*{1.0cm}     & 0.6        \hspace*{0.8cm} & 0.6        \hspace*{0.8cm} & 0.6        \hspace*{0.8cm} & 0.6         \hspace*{1.0cm}\\%
Young stars per pixel\dotfill       & 1      \hspace*{1.0cm}     & 480        \hspace*{0.8cm} & 200        \hspace*{0.8cm} & 571        \hspace*{0.8cm} & 238        \hspace*{1.0cm} \\%
Young population M$_{up}$\dotfill   & $-$1.1 \hspace*{1.0cm}     & $-$1.1     \hspace*{0.8cm} & $-$1.1     \hspace*{0.8cm} & $-$2.2     \hspace*{0.8cm} & $-$2.2      \hspace*{1.0cm}\\%
Young population M$_{low}$\dotfill  & 15.5   \hspace*{1.0cm}     & 15.5       \hspace*{0.8cm} & 15.5       \hspace*{0.8cm} & 15.5       \hspace*{0.8cm} & 15.5        \hspace*{1.0cm}\\%
$\mu_V$ (mag)\dotfill               & 21.56  \hspace*{1.0cm}     & 21.38      \hspace*{0.8cm} & 22.02      \hspace*{0.8cm} & 21.20      \hspace*{0.8cm} & 21.88       \hspace*{1.0cm}\\%
$\sigma$ (mag)\dotfill              & 0.41 (37\%)                & 0.08 (7\%)                 & 0.07 (7\%)                 & 0.10 (9\%)                 & 0.13 (12\%) \\%

\enddata
\tablecomments{$\mu_V$ is the mean magnitude produced in each simulation; $\sigma$ is the pixel-to-pixel
variations around that mean, expressed in magnitudes and percentage of the mean.}
\label{tabvir}
\end{deluxetable}
\end{landscape}

%% file: Gonzalez.tab4.tex
\begin{deluxetable}{rrrrr}
\tablewidth{3in}

\tablecaption{LMC Stellar luminosity function}
\tablehead{
\colhead{M$_V$} & 
\colhead{$\Phi$(M$_V$)} &
\colhead{} &
\colhead{M$_V$} & 
\colhead{$\Phi$(M$_V$)} \\%
\\[-0.3cm]
}
\startdata

-1  &   1  & ~~~~~~~~ &  8 & 180 \\%
 0  &   4  & &  9 & 224 \\%
 1  &  11  & & 10 & 376 \\%
 2  &  16  & & 11 & 480 \\%
 3  &  41  & & 12 & 683 \\%
 4  &  57  & & 13 & 546 \\%
 5  &  98  & & 14 & 546 \\%
 6  & 100  & & 15 & 683 \\%
 7  & 157  & &    &     \\%
\enddata
\tablecomments{Stellar luminosity function $\Phi$(M$_V$) used for the LMC. 
Units are the same as in \cite{jahr97} for the solar neighborhood (number of 
stars within 20 pc in 1 mag bin). 
The mean $V$ surface brightness of the ``mother'' LMC simulation is  
$\mu_V$ = 21.68; the spread around the mean has a FWHM
of 0.29 mag (26\% of the mean); the pixel-to-pixel variation around the mean
is $\sigma$ = 4.79 mag (441\% of the mean).}
\label{tphi}
\end{deluxetable}

%% file: Gonzalez.tab5.tex
\begin{deluxetable}{ccc}
\tablewidth{3in}

\tablecaption{Parameters of the radial M~31 simulations}
\tablehead{
\colhead{Radius ($'$)} & 
\colhead{$\mu_V$ (mag)} &
\colhead{Stars per pixel} \\[-0.3cm]
}
\startdata
 10 & 20.3 & 165 \\%
 20 & 21.0 &  85 \\%
 30 & 21.6 &  50 \\%
 40 & 21.9 &  38 \\%
 50 & 22.1 &  31 \\%
 60 & 22.7 &  18 \\%
 70 & 23.2 &  11.5 \\%
 80 & 23.8 &   9.5 \\%
 90 & 24.5 &   3.5 \\%
100 & 25.1 &   2.0 \\%

\enddata
\tablecomments{$\mu_V$ is the mean V surface brightness in magnitudes 
from \citet{walt88}.}
\label{tabm31}
\end{deluxetable}

%% file: Gonzalez.tab6.tex
{

\begin{landscape}
\begin{deluxetable}{lllllllllll}
\tablewidth{9.3in}

\tablecaption{Parameters of the simulations}
\tablehead{
\colhead{Field} &
\colhead{Pixel} &
\colhead{Tile} &
\colhead{Stars} &
\colhead{PSF} &
\colhead{backg.} &
\colhead{exp.\ time} &
\colhead{gain} &
\colhead{RON} &
\colhead{N$_{gal}$} &
\colhead{N$_{ref}$} \\
\colhead{} &
\colhead{size ($^{\prime\prime}$)} &
\colhead{size (pixel)} &
\colhead{per pixel} &
\colhead{} &
\colhead{(mag)} &
\colhead{(s)} &
\colhead{} &
\colhead{(e-)} &
\colhead{} &
\colhead{} 
}
\startdata 
M~31--Inner WFPC2               & 0.1    & 100   & 50            & WFC        & 23.01 & $\infty$ &            &      & 18$\pm$15    & 21$\pm$16 (2.5e3 s) \\%
M~31--Inner WFPC2 2-hr          & 0.1    & 100   & 50            & WFC        & 23.01 & 7200     & 7          & 10.4 & 18$\pm$15    &                     \\%
M~31--Inner ACS--WFC            & 0.05   & 200   & 12.5          & ACS--WFC   & 23.01 & $\infty$ &            &      & 38$\pm$22    &                     \\%
M~31--Inner ACS--WFC 2-hr       & 0.05   & 200   & 12.5          & ACS--WFC   & 23.01 & 7200     & 3$\times$8 & 7.1  & 38$\pm$22    &                     \\%
M~31--Inner ACS--HR             & 0.026  & 400   & 3.4           & ACS--HR    & 23.01 & $\infty$ &            &      & 36$\pm$21    &                     \\%
M~31--Inner ACS--HR 2-hr        & 0.026  & 400   & 3.4           & ACS--HR    & 23.01 & 7200     & 3$\times$8 & 5.9  & 36$\pm$21    &                     \\%
M~31--Inner 3-m real            & 0.02   & 500   & 2             & ACS--HR    & 23.01 & $\infty$ &            &      & 35$\pm$21    &                     \\%
M~31--Inner 3-m real 2-hr       & 0.02   & 500   & 2             & ACS--H     & 23.01 & 7200     & 7          & 10.4 & 35$\pm$21    &                     \\%
M~31--Inner 4.5-m real          & 0.014  & 700   & 1             & ACS--HR    & 23.01 & $\infty$ &            &      & 43$\pm$23    &                     \\%
M~31--Inner 4.5-m real 2-hr     & 0.014  & 700   & 1             & ACS--HR    & 23.01 & 7200     & 7          & 10.4 & 43$\pm$23    &                     \\%
M~31--Inner 6-m real            & 0.01   & 1000  & 0.5           & ACS--HR    & 23.01 & $\infty$ &            &      & 69$\pm$29    &                     \\%
M~31--Inner 6-m real 2-hr       & 0.01   & 1000  & 0.5           & ACS--HR    & 23.01 & 7200     & 7          & 10.4 & 43$\pm$23    &                     \\%
M~31--Inner 10-m real           & 0.006  & 1600  & 0.2           & ACS--HR    & 23.01 & $\infty$ &            &      & 139$\pm$41   &                     \\%
M~31--Inner 10-m real 2-hr      & 0.006  & 1600  & 0.2           & ACS--HR    & 23.01 & 7200     & 7          & 10.4 & 99$\pm$35    &                     \\%
M~31--Inner 0.8-m gauss         & 0.08   & 100   & 32            & Gaussian   & 23.01 & $\infty$ &            &      & 19$\pm$15    &                     \\%
M~31--Inner 0.8-m gauss 2-hr    & 0.08   & 100   & 32            & Gaussian   & 23.01 & 7200     & 7          & 10.4 & 19$\pm$15    &                     \\%
M~31--Inner 1-m gauss           & 0.06   & 100   & 16            & Gaussian   & 23.01 & $\infty$ &            &      & 28$\pm$19    &                     \\%
M~31--Inner 1-m gauss 2-hr      & 0.06   & 100   & 16            & Gaussian   & 23.01 & 7200     & 7          & 10.4 & 28$\pm$19    &                     \\%
M~31--Inner 1.5-m gauss         & 0.04   & 100   & 8             & Gaussian   & 23.01 & $\infty$ &            &      & 42$\pm$23    &                     \\%
M~31--Inner 1.5-m gauss 2-hr    & 0.04   & 100   & 8             & Gaussian   & 23.01 & 7200     & 7          & 10.4 & 42$\pm$23    &                     \\%
M~31--Inner 2-m gauss           & 0.03   & 100   & 4             & Gaussian   & 23.01 & $\infty$ &            &      & 53$\pm$25    &                     \\%
M~31--Inner 2-m gauss 2-hr      & 0.03   & 100   & 4             & Gaussian   & 23.01 & 7200     & 7          & 10.4 & 53$\pm$25    &                     \\%
M~31--Inner 3-m gauss           & 0.02   & 100   & 2             & Gaussian   & 23.01 & $\infty$ &            &      & 71$\pm$29    &                     \\%
M~31--Inner 3-m gauss 2-hr      & 0.02   & 100   & 2             & Gaussian   & 23.01 & 7200     & 7          & 10.4 & 71$\pm$29    &                     \\%
M~31--Inner 4.5-m gauss         & 0.014  & 100   & 1             & Gaussian   & 23.01 & $\infty$ &            &      & 172$\pm$46   &                     \\%
M~31--Inner 4.5-m gauss 2-hr    & 0.014  & 100   & 1             & Gaussian   & 23.01 & 7200     & 7          & 10.4 & 172$\pm$46   &                     \\%
M~31--Inner 6-m gauss           & 0.01   & 100   & 0.5           & Gaussian   & 23.01 & $\infty$ &            &      & 211$\pm$51   &                     \\%
M~31--Inner 6-m gauss 2-hr      & 0.01   & 100   & 0.5           & Gaussian   & 23.01 & 7200     & 7          & 10.4 & 181$\pm$47   &                     \\%
M~31--Inner 10-m gauss          & 0.006  & 100   & 0.2           & Gaussian   & 23.01 & $\infty$ &            &      & 582$\pm$84   &                     \\%
M~31--Inner 10-m gauss 2-hr     & 0.006  & 100   & 0.2           & Gaussian   & 23.01 & 7200     & 7          & 10.4 & 255$\pm$56   &                     \\%
M~31 10'                        & 0.1    & 100   & 165           & WFC        & 23.01 & $\infty$ &            &      & 3$\pm$6      &                     \\%
M~31 10' 0.5-hr                 & 0.1    & 100   & 165           & WFC        & 23.01 & 1800     & 7          & 7.4  & 3$\pm$6      &                     \\%
M~31 20'                        & 0.1    & 100   & 85            & WFC        & 23.01 & $\infty$ &            &      & 10$\pm$11    &                     \\%
M~31 20' 0.5-hr                 & 0.1    & 100   & 85            & WFC        & 23.01 & 1800     & 7          & 7.4  & 10$\pm$11    &                     \\%
M~31 30'                        & 0.1    & 100   & 50            & WFC        & 23.01 & $\infty$ &            &      & 18$\pm$15    & 21$\pm$16 (2.5e3 s) \\%
M~31 30' 0.5-hr                 & 0.1    & 100   & 50            & WFC        & 23.01 & 1800     & 7          & 7.4  & 18$\pm$15    &                     \\%
M~31 40'                        & 0.1    & 100   & 38            & WFC        & 23.01 & $\infty$ &            &      & 20$\pm$16    &                     \\%
M~31 40' 0.5-hr                 & 0.1    & 100   & 38            & WFC        & 23.01 & 1800     & 7          & 7.4  & 16$\pm$14    &                     \\%
M~31 50'                        & 0.1    & 100   & 31            & WFC        & 23.01 & $\infty$ &            &      & 20$\pm$16    &                     \\%
M~31 50' 0.5-hr                 & 0.1    & 100   & 31            & WFC        & 23.01 & 1800     & 7          & 7.4  & 16$\pm$14    &                     \\%
M~31 60'                        & 0.1    & 100   & 18            & WFC        & 23.01 & $\infty$ &            &      & 36$\pm$21    &                     \\%
M~31 60' 0.5-hr                 & 0.1    & 100   & 18            & WFC        & 23.01 & 1800     & 7          & 7.4  & 30$\pm$19    &                     \\%
M~31 70'                        & 0.1    & 100   & 11.5          & WFC        & 23.01 & $\infty$ &            &      & 53$\pm$25    &                     \\%
M~31 70' 0.5-hr                 & 0.1    & 100   & 11.5          & WFC        & 23.01 & 1800     & 7          & 7.4  & 43$\pm$23    &                     \\%
M~31 80'                        & 0.1    & 100   & 9.5           & WFC        & 23.01 & $\infty$ &            &      & 43$\pm$23    &                     \\%
M~31 80' 0.5-hr                 & 0.1    & 100   & 9.5           & WFC        & 23.01 & 1800     & 7          & 7.4  & 33$\pm$20    &                     \\%
M~31 90'                        & 0.1    & 100   & 3.5           & WFC        & 23.01 & $\infty$ &            &      & 96$\pm$34    &                     \\%
M~31 90' 0.5-hr                 & 0.1    & 100   & 3.5           & WFC        & 23.01 & 1800     & 7          & 7.4  & 66$\pm$28    &                     \\%
M~31 100'                       & 0.1    & 100   & 2             & WFC        & 23.01 & $\infty$ &            &      & 99$\pm$35    &                     \\%
M~31 100' 0.5-hr                & 0.1    & 100   & 2             & WFC        & 23.01 & 1800     & 7          & 7.4  & 73$\pm$30    &                     \\%
LMC WFPC2                       & 0.1    & 800   & 0.027         & WFC        & 23.25 & $\infty$ &            &      & 83$\pm$32    & 66$\pm$28 (1000 s)  \\%
LMC WFPC2 0.5-hr                & 0.1    & 800   & 0.027         & WFC        & 23.25 & 1800     & 7          & 7.4  & 46$\pm$24    &                     \\%
LMC ACS--WFC                    & 0.05   & 1520  & 7 10$^{-3}$   & ACS--WFC   & 23.25 & $\infty$ &            &      & 92$\pm$34    &                     \\%
LMC ACS--WFC 0.5-hr             & 0.05   & 1520  & 7 10$^{-3 }$  & ACS--WFC   & 23.25 & 1800     & 3$\times$7 & 7.1  & 81$\pm$32    &                     \\%
LMC ACS--HR                     & 0.026  & 2815  & 2 10$^{-3}$   & ACS--HR    & 23.25 & $\infty$ &            &      & 145$\pm$42   &                     \\%
LMC ACS--HR  0.5-hr             & 0.026  & 2815  & 2 10$^{-3}$   & ACS--HR    & 23.25 & 1800     & 3$\times$7 & 5.9  & 92$\pm$34    &                     \\%
LMC 3-m real                    & 0.02   & 3799  & 1.1 10$^{-3}$ & ACS--HR    & 23.25 & $\infty$ &            &      & 119$\pm$38   &                     \\%
LMC 3-m real 0.5-hr             & 0.02   & 3799  & 1.1 10$^{-3}$ & ACS--HR    & 23.25 & 1800     & 7          & 7.4  & 63$\pm$28    &                     \\%
LMC 4.5-m real                  & 0.014  & 5377  & 5 10$^{-4 }$  & ACS--HR    & 23.25 & $\infty$ &            &      & 107$\pm$36   &                     \\%
LMC 4.5-m real 0.5-hr           & 0.014  & 5377  & 5 10$^{-4 }$  & ACS--HR    & 23.25 & 1800     & 7          & 7.4  & 64$\pm$28    &                     \\%
LMC 10-m real                   & 0.006  & 12040 & 1 10$^{-4}$   & ACS--HR    & 23.25 & $\infty$ &            &      & 251$\pm$55   &                     \\%
LMC 10-m real 0.5-hr            & 0.006  & 12040 & 1 10$^{-4}$   & ACS--HR    & 23.25 & 1800     & 7          & 7.4  & 112$\pm$37   &                     \\%
LMC 0.8-m gauss                 & 0.08   & 950   & 0.017         & Gaussian   & 23.25 & $\infty$ &            &      & 124$\pm$39   &                     \\%
LMC 0.8-m gauss 0.5-hr          & 0.08   & 950   & 0.017         & Gaussian   & 23.25 & 1800     & 7          & 7.4  & 54$\pm$26    &                     \\%
LMC 1-m gauss                   & 0.06   & 1343  & 8.5 10$^{-3}$ & Gaussian   & 23.25 & $\infty$ &            &      & 231$\pm$53   &                     \\%
LMC 1-m gauss 0.5-hr            & 0.06   & 1343  & 8.5 10$^{-3}$ & Gaussian   & 23.25 & 1800     & 7          & 7.4  & 74$\pm$30    &                     \\%
LMC 1.5-m gauss                 & 0.04   & 1900  & 4.3 10$^{-3}$ & Gaussian   & 23.25 & $\infty$ &            &      & 395$\pm$70   &                     \\%
LMC 1.5-m gauss 0.5-hr          & 0.04   & 1900  & 4.3 10$^{-3}$ & Gaussian   & 23.25 & 1800     & 7          & 7.4  & 94$\pm$34    &                     \\%
LMC 2-m gauss                   & 0.03   & 2688  & 2.1 10$^{-3}$ & Gaussian   & 23.25 & $\infty$ &            &      & 789$\pm$98   &                     \\%
LMC 2-m gauss 0.5-hr            & 0.03   & 2688  & 2.1 10$^{-3}$ & Gaussian   & 23.25 & 1800     & 7          & 7.4  & 191$\pm$48   &                     \\%
LMC 3-m gauss                   & 0.02   & 3799  & 1.1 10$^{-3}$ & Gaussian   & 23.25 & $\infty$ &            &      & 1084$\pm$115 &                     \\%
LMC 3-m gauss 0.5-hr            & 0.02   & 3799  & 1.1 10$^{-3}$ & Gaussian   & 23.25 & 1800     & 7          & 7.4  & 381$\pm$68   &                     \\%
LMC 4.5-m gauss                 & 0.014  & 5377  & 5 10$^{-4}$   & Gaussian   & 23.25 & $\infty$ &            &      & 1673$\pm$143 &                     \\%
LMC 4.5-m gauss 0.5-hr          & 0.014  & 5377  & 5 10$^{-4}$   & Gaussian   & 23.25 & 1800     & 7          & 7.4  & 455$\pm$75   &                     \\%
LMC 10-m gauss                  & 0.006  & 12040 & 1 10$^{-4}$   & Gaussian   & 23.25 & $\infty$ &            &      & 1923$\pm$153 &                     \\%
LMC 10-m gauss  0.5-hr          & 0.006  & 12040 & 1 10$^{-4 }$  & Gaussian   & 23.25 & 1800     & 7          & 7.4  & 522$\pm$80   &                     \\%
Virgo IA incl WFPC2             & 0.1    & 100 & 2.4 10$^{4}$    & WFC        & 23.01 & $\infty$ &            &      & 201$\pm$50   & 114$\pm$37\tablenotemark{a}\ (2 10$^{4}$ s)\\%
Virgo IA incl WFPC2 2-hr        & 0.1    & 100 & 2.4 10$^{4}$    & WFC        & 23.01 & 7200     & 7          & 10.4 & 201$\pm$50   &                     \\%
Virgo IA incl ACS--WFC          & 0.05   & 100 & 6 10$^{3}$      & ACS--WFC   & 23.01 & $\infty$ &            &      & 271$\pm$58   &                     \\%
Virgo IA incl ACS--WFC 2-hr     & 0.05   & 100 & 6 10$^{3}$      & ACS--WFC   & 23.01 & 7200     & 3$\times$8 & 7.1  & 271$\pm$58   &                     \\%
Virgo IA incl ACS--HR           & 0.026  & 100 & 1.6 10$^{3}$    & ACS--HR    & 23.01 & $\infty$ &            &      & 185$\pm$48   &                     \\%
Virgo IA incl ACS--HR 2-hr      & 0.026  & 100 & 1.6 10$^{3}$    & ACS--HR    & 23.01 & 7200     & 3$\times$8 & 5.9  & 185$\pm$48   &                     \\%
Virgo IA incl 4.5-m real        & 0.014  & 100 & 480             & ACS--HR    & 23.01 & $\infty$ &            &      & 248$\pm$55   &                     \\%
Virgo IA incl 4.5-m real 2-hr   & 0.014  & 100 & 480             & ACS--HR    & 23.01 & 7200     & 7          & 10.4 & 208$\pm$50   &                     \\%
Virgo IA incl 10-m real         & 0.006  & 100 & 96              & ACS--HR    & 23.01 & $\infty$ &            &      & 373$\pm$68   &                     \\%
Virgo IA incl 10-m real 2-hr    & 0.006  & 100 & 96              & ACS--HR    & 23.01 & 7200     & 7          & 10.4 & 224$\pm$52   &                     \\%
Virgo IA incl 1.2-m gauss       & 0.05   & 100 & 6 10$^{3}$      & Gaussian   & 23.01 & $\infty$ &            &      & 172$\pm$46   &                     \\%
Virgo IA incl 1.2-m gauss 2-hr  & 0.05   & 100 & 6 10$^{3}$      & Gaussian   & 23.01 & 7200     & 7          & 10.4 & 172$\pm$46   &                     \\%
Virgo IA incl 2.4-m gauss       & 0.026  & 100 & 1.6 10$^{3}$    & Gaussian   & 23.01 & $\infty$ &            &      & 168$\pm$45   &                     \\%
Virgo IA incl 2.4-m gauss 2-hr  & 0.026  & 100 & 1.6 10$^{3}$    & Gaussian   & 23.01 & 7200     & 7          & 10.4 & 145$\pm$42   &                     \\%
Virgo IA incl 4.5-m gauss       & 0.014  & 100 & 480             & Gaussian   & 23.01 & $\infty$ &            &      & 300$\pm$61   &                     \\%
Virgo IA incl 4.5-m gauss 2-hr  & 0.014  & 100 & 480             & Gaussian   & 23.01 & 7200     & 7          & 10.4 & 231$\pm$53   &                     \\%
Virgo IA incl 10-m gauss        & 0.006  & 100 & 96              & Gaussian   & 23.01 & $\infty$ &            &      & 399$\pm$70   &                     \\%
Virgo IA incl 10-m gauss 2-hr   & 0.006  & 100 & 96              & Gaussian   & 23.01 & 7200     & 7          & 10.4 & 248$\pm$55   &                     \\%
Virgo arm incl WFPC2            & 0.1    & 100 & 2.4 10$^{4}$    & WFC        & 23.01 & $\infty$ &            &      & 117$\pm$38   & 84$\pm$32 (2 10$^{4}$ s) \\%
Virgo arm incl WFPC2 2-hr       & 0.1    & 100 & 2.4 10$^{4}$    & WFC        & 23.01 & 7200     & 7          & 10.4 & 117$\pm$38   &                     \\%
Virgo arm incl ACS--WFC         & 0.05   & 100 & 6 10$^{3}$      & ACS--WFC   & 23.01 & $\infty$ &            &      & 140$\pm$41   &                     \\%
Virgo arm incl ACS--WFC 2-hr    & 0.05   & 100 & 6 10$^{3}$      & ACS--WFC   & 23.01 & 7200     & 3$\times$8 & 7.1  & 140$\pm$41   &                     \\%
Virgo arm incl ACS--HR          & 0.026  & 100 & 1.6 10$^{3}$    & ACS--HR    & 23.01 & $\infty$ &            &      & 116$\pm$38   &                     \\%
Virgo arm incl ACS--HR 2-hr     & 0.026  & 100 & 1.6 10$^{3}$    & ACS--HR    & 23.01 & 7200     & 3$\times$8 & 5.9  & 106$\pm$36   &                     \\%
Virgo arm incl 4.5-m real       & 0.014  & 100 & 480             & ACS--HR    & 23.01 & $\infty$ &            &      & 175$\pm$46   &                     \\%
Virgo arm incl 4.5-m real 2-hr  & 0.014  & 100 & 480             & ACS--HR    & 23.01 & 7200     & 7          & 10.4 & 142$\pm$42   &                     \\%
Virgo arm incl 10-m real        & 0.006  & 100 & 96              & ACS--HR    & 23.01 & $\infty$ &            &      & 221$\pm$52   &                     \\%
Virgo arm incl 10-m real 2-hr   & 0.006  & 100 & 96              & ACS--HR    & 23.01 & 7200     & 7          & 10.4 & 191$\pm$48   &                     \\%
Virgo arm incl 1.2-m gauss      & 0.05   & 100 & 6 10$^{3}$      & Gaussian   & 23.01 & $\infty$ &            &      & 106$\pm$36   &                     \\%
Virgo arm incl 1.2-m gauss 2-hr & 0.05   & 100 & 6 10$^{3}$      & Gaussian   & 23.01 & 7200     & 7          & 10.4 & 106$\pm$36   &                     \\%
Virgo arm incl 2.4-m gauss      & 0.026  & 100 & 1.6 10$^{3}$    & Gaussian   & 23.01 & $\infty$ &            &      & 116$\pm$38   &                     \\%
Virgo arm incl 2.4-m gauss 2-hr & 0.026  & 100 & 1.6 10$^{3}$    & Gaussian   & 23.01 & 7200     & 7          & 10.4 &  92$\pm$34   &                     \\%
Virgo arm incl 4.5-m gauss      & 0.014  & 100 & 480             & Gaussian   & 23.01 & $\infty$ &            &      & 188$\pm$48   &                     \\%
Virgo arm incl 4.5-m gauss 2-hr & 0.014  & 100 & 480             & Gaussian   & 23.01 & 7200     & 7          & 10.4 & 162$\pm$45   &                     \\%
Virgo arm incl 10-m gauss       & 0.006  & 100 & 96              & Gaussian   & 23.01 & $\infty$ &            &      & 234$\pm$54   &                     \\%
Virgo arm incl 10-m gauss 2-hr  & 0.006  & 100 & 96              & Gaussian   & 23.01 & 7200     & 7          & 10.4 & 211$\pm$51   &                     \\%
Virgo IA FO WFPC2               & 0.1    & 100 & 1 10$^{4}$      & WFC        & 23.01 & $\infty$ &            &      & 256$\pm$56   &                     \\%
Virgo IA FO WFPC2 2-hr          & 0.1    & 100 & 1 10$^{4}$      & WFC        & 23.01 & 7200     & 7          & 10.4 & 256$\pm$56   &                     \\%
Virgo IA FO ACS--WFC            & 0.05   & 100 & 2.5 10$^{3}$    & ACS--WFC   & 23.01 & $\infty$ &            &      & 419$\pm$72   &                     \\%
Virgo IA FO ACS--WFC 2-hr       & 0.05   & 100 & 2.5 10$^{3}$    & ACS--WFC   & 23.01 & 7200     & 3$\times$8 & 7.1  & 419$\pm$72   &                     \\%
Virgo IA FO ACS--HR             & 0.026  & 100 & 720             & ACS--HR    & 23.01 & $\infty$ &            &      & 356$\pm$66   &                     \\%
Virgo IA FO ACS--HR 2-hr        & 0.026  & 100 & 720             & ACS--HR    & 23.01 & 7200     & 3$\times$8 & 5.9  & 317$\pm$62   &                     \\%
Virgo IA FO 4.5-m real          & 0.014  & 100 & 204             & ACS--HR    & 23.01 & $\infty$ &            &      & 505$\pm$79   &                     \\%
Virgo IA FO 4.5-m real 2-hr     & 0.014  & 100 & 204             & ACS--HR    & 23.01 & 7200     & 7          & 10.4 & 370$\pm$67   &                     \\%
Virgo IA FO 10-m real           & 0.006  & 100 & 40              & ACS--HR    & 23.01 & $\infty$ &            &      & 627$\pm$88   &                     \\%
Virgo IA FO 10-m real 2-hr      & 0.006  & 100 & 40              & ACS--HR    & 23.01 & 7200     & 7          & 10.4 & 343$\pm$65   &                     \\%
Virgo IA FO 1.2-m gauss         & 0.05   & 100 & 2.5 10$^{3}$    & Gaussian   & 23.01 & $\infty$ &            &      & 248$\pm$55   &                     \\%
Virgo IA FO 1.2-m gauss 2-hr    & 0.05   & 100 & 2.5 10$^{3}$    & Gaussian   & 23.01 & 7200     & 7          & 10.4 & 248$\pm$55   &                     \\%
Virgo IA FO 2.4-m gauss         & 0.026  & 100 & 720             & Gaussian   & 23.01 & $\infty$ &            &      & 333$\pm$64   &                     \\%
Virgo IA FO 2.4-m gauss 2-hr    & 0.026  & 100 & 720             & Gaussian   & 23.01 & 7200     & 7          & 10.4 & 238$\pm$54   &                     \\%
Virgo IA FO 4.5-m gauss         & 0.014  & 100 & 204             & Gaussian   & 23.01 & $\infty$ &            &      & 515$\pm$79   &                     \\%
Virgo IA FO 4.5-m gauss 2-hr    & 0.014  & 100 & 204             & Gaussian   & 23.01 & 7200     & 7          & 10.4 & 350$\pm$65   &                     \\%
Virgo IA FO 10-m gauss          & 0.006  & 100 & 40              & Gaussian   & 23.01 & $\infty$ &            &      & 700$\pm$93   &                     \\%
Virgo IA FO 10-m gauss 2-hr     & 0.006  & 100 & 40              & Gaussian   & 23.01 & 7200     & 7          & 10.4 & 396$\pm$79   &                     \\%
Virgo arm FO WFPC2              & 0.1    & 100 & 1 10$^{4}$      & WFC        & 23.01 & $\infty$ &            &      & 188$\pm$48   &                     \\%
Virgo arm FO WFPC2 2-hr         & 0.1    & 100 & 1 10$^{4}$      & WFC        & 23.01 & 7200     & 7          & 10.4 & 188$\pm$48   &                     \\%
Virgo arm FO ACS--WFC           & 0.05   & 100 & 2.5 10$^{3}$    & ACS--WFC   & 23.01 & $\infty$ &            &      & 257$\pm$56   &                     \\%
Virgo arm FO ACS--WFC 2-hr      & 0.05   & 100 & 2.5 10$^{3}$    & ACS--WFC   & 23.01 & 7200     & 3$\times$8 & 7.1  & 257$\pm$56   &                     \\%
Virgo arm FO ACS--HR            & 0.026  & 100 & 720             & ACS--HR    & 23.01 & $\infty$ &            &      & 290$\pm$60   &                     \\%
Virgo arm FO ACS--HR 2-hr       & 0.026  & 100 & 720             & ACS--HR    & 23.01 & 7200     & 3$\times$8 & 5.9  & 254$\pm$56   &                     \\%
Virgo arm FO 4.5-m real         & 0.014  & 100 & 204             & ACS--HR    & 23.01 & $\infty$ &            &      & 330$\pm$64   &                     \\%
Virgo arm FO 4.5-m real 2-hr    & 0.014  & 100 & 204             & ACS--HR    & 23.01 & 7200     & 7          & 10.4 & 228$\pm$53   &                     \\%
Virgo arm FO 10-m real          & 0.006  & 100 & 40              & ACS--HR    & 23.01 & $\infty$ &            &      & 452$\pm$74   &                     \\%
Virgo arm FO 10-m real 2-hr     & 0.006  & 100 & 40              & ACS--HR    & 23.01 & 7200     & 7          & 10.4 & 284$\pm$59   &                     \\%
Virgo arm FO 1.2-m gauss        & 0.05   & 100 & 2.5 10$^{3}$    & Gaussian   & 23.01 & $\infty$ &            &      & 175$\pm$46   &                     \\%
Virgo arm FO 1.2-m gauss 2-hr   &  0.05  & 100 & 2.5 10$^{3}$    & Gaussian   & 23.01 & 7200     & 7          & 10.4 & 175$\pm$46   &                     \\%
Virgo arm FO 2.4-m gauss        & 0.026  & 100 & 720             & Gaussian   & 23.01 & $\infty$ &            &      & 310$\pm$62   &                     \\%
Virgo arm FO 2.4-m gauss 2-hr   & 0.026  & 100 & 720             & Gaussian   & 23.01 & 7200     & 7          & 10.4 & 224$\pm$52   &                     \\%
Virgo arm FO 4.5-m gauss        & 0.014  & 100 & 204             & Gaussian   & 23.01 & $\infty$ &            &      & 376$\pm$68   &                     \\%
Virgo arm FO 4.5-m gauss 2-hr   & 0.014  & 100 & 204             & Gaussian   & 23.01 & 7200     & 7          & 10.4 & 254$\pm$56   &                     \\%
Virgo arm FO 10-m gauss         & 0.006  & 100 & 40              & Gaussian   & 23.01 & $\infty$ &            &      & 452$\pm$74   &                     \\%
Virgo arm FO 10-m gauss 2-hr    & 0.006  & 100 & 40              & Gaussian   & 23.01 & 7200     & 7          & 10.4 & 315$\pm$62   &                     \\%
\enddata

\tablecomments{
Col.\ (1) In Virgo, In Virgo, simulations noted ``incl'' correspond to those for an inclination of 65\degr, while those
noted ``FO'' correspond to face--on. Col.\ (2) Tile size in linear number of pixels on the side. Col.\ (4)  
The number of stars per pixel is for the old stellar population, except for the LMC, where it refers to total 
stars. Col.\ (10) Number of galaxies in a field 5.3 arcmin$^2$. Col.\ (11) Number of galaxies in a field 5.3 arcmin$^2$; exposure time of data included in parentheses.} 
\tablenotetext{a}{Image contains some ``arm.'' The galaxy number--counts, however, are still compatible with the
simulation, within the quoted errors.}
\label{tsim2}
\end{deluxetable}
\end{landscape}

}